\documentstyle[12pt]{article}

0

\begin{document}

\newcommand{\MqV}{$\cal M$$_{q}$($\cal V$)}
\def\II{\relax{\rm 1\kern-.35em1}}
\def\IP{\relax{\rm I\kern-.18em P}}
\renewcommand{\theequation}{\thesection.\arabic{equation}}
\csname @addtoreset\endcsname{equation}{section}

\hspace{12cm} hep-th/9604043

\begin{center}

{}~\vfill

{\large \bf  Integrability, Duality and Strings$^{\dagger}$}

\end{center}

\vspace{20 mm}

\begin{center}

{\bf C\'{e}sar G\'{o}mez$^a$, Rafael Hern\'{a}ndez$^{a,b}$ and
Esperanza L\'{o}pez$^c$} 
  
\vspace{10 mm}

$^a${\em Instituto de Matem\'{a}ticas y F\'{\i}sica Fundamental,
CSIC, \protect \\ Serrano 123, 28006 Madrid, Spain}  
  
\vspace{6 mm}   
  
$^b${\em Departamento de F\'{\i}sica Te\'{o}rica, C-XI,
Universidad Aut\'{o}noma de Madrid, \protect \\
Cantoblanco, 28049 Madrid, Spain}
  
\vspace{6 mm}
  
$^c${\em Institute for Theoretical Physics, University of
California, \protect \\ Santa Barbara, CA 93106-4030}

\end{center}      

\vspace{6cm}

$^{\dagger}$ \hspace{2 mm} Lecture given by C.G. at the
``Quantum Field Theory Workshop'', August 1996, Bulgary.    

\pagebreak


\section{Introduction.}

After the two years ago work of Seiberg and Witten \cite{SW,SW2}, the 
Pandora box of string theory has been opened once more. The magic opening 
word has been {\em duality\/}. A web of exciting interrelated results 
it contains has appeared during the last months: string-string duality 
($U$ duality) \cite{HT}; the physical interpretation of the conifold 
singularity \cite{bh}; heterotic-type II dual pairs \cite{HT,svd,r1,r2,KV}; 
R-R states 
and Dirichlet-branes \cite{P,PW}; derivation from $M$-theory of  
dualities in string theory; and non-perturbative enhancement of 
symmetries \cite{KMP,KM,BSV}. 
  
String-string duality was first pointed out between heterotic strings 
compactified 
on $T^4$, and type II$_A$ strings on $K3$ \cite{svd}. The first check of this 
duality relation between these two different approaches to string theory 
is of course trying to understand the equivalence, for the type II$_A$ 
string on $K3$, of the well known enhancement of symmetry for the heterotic 
string on $T^4$ at certain points in the moduli space. This problem was 
beatifully solved in \cite{svd}, where the enhanced non abelian gauge 
symmetries 
appear associated to the orbifold singularities of $K3$. These singularities 
are of $A$-$D$-$E$ type, and its combinaton corresponds to a group of total 
rank $\leq 20$. Geometrically, these singularities arise from the collapse 
of a set of $2$-cycles; the number of collapsing cycles equals the rank of the 
gauge group, and the intersection matrix is given by the Dynkin diagram of 
the singularity. This geometry can be directly connected with Strominger's 
suggestion for the interpretation, in Calabi-Yau threefolds, of conifold 
singularities, where a $3$-cycle collapses to a point; in this case, and 
for the 
type II$_B$, a massless soliton with R-R charge can be interpreted in terms 
of an appropiated $3$-brane wrapping around the $3$-cycle. For the case of 
$K3$, the enhancement of symmetry is interpreted in terms of $2$-branes 
wrapping around $2$-cycles that collapse at the orbifold point.
  
String-string duality in six dimensions can now be used to produce dual 
heterotic-type II pairs in four dimensions, by compactifying on a 
$2$-torus. This 
is the point where string-string duality and type II $T$-duality produces 
the desired $S$ duality for the heterotic string \cite{DMW}.
  
The dual pairs become then the string analog of the celebrated 
Seiberg-Witten solution of $N\!=\!2$ gauge theories. In fact, in the 
simpler case of supersymmetric gauge theories, the quantum moduli of a 
particular 
theory with some gauge group and matter content is characterized by the 
classical special geometry of the moduli of a curve, while in the string 
dual pairs context the quantum moduli of a certain heterotic string 
compactified 
on some Calabi-Yau manifold $X$ is given by the classical special geometry 
of the complex structures moduli space of some other Calabi-Yau manifold $Y$, 
in which the type II string theory is compactified.
  
Taking now into account that field theory is the point particle limit of 
string theory, it was very natural to expect that the Seiberg-Witten 
quantum moduli 
for $N\!=\!2$ gauge theories should be naturally derived as the point particle 
limit of heterotic-type II dual pairs, provided the low energy theory defined 
by the heterotic string corresponds to the desired $N\!=\!2$
gauge theory \cite{KKLMV,GLi}. 
This is in fact what happens when a singular point in the moduli space of the 
type II theory is blown up; this point is in the weak coupling and point 
particle 
limit region of the moduli, and is a non abelian enhancement of symmetry 
point, with the non abelian symmetry that of the $N\!=\!2$ gauge theory under 
study.
  
Already at this level of the discussion an interesting subtlety, connecting 
a non trivial way global symmetries of field theory and 
string theory global symmetries, appears. In fact, the point particle limit 
of the string defines 
a coordinate for the field theory quantum moduli which involves both the 
string tension and the heterotic dilaton, both taken in a double scaling 
limit. The importance of this fact is that for the corresponding moduli of the 
dual type II string we can define global stringy transformations, associated 
to symmetries of the Calabi-Yau threefold $Y$, which {\em induce\/} 
transformations on the field theory quantum moduli parameter. These 
transformations have, at the field theory level, the interpretation of 
global $R$-symmetries that we can not quotient by.
  
The above described mechanism, for the enhancement of non abelian 
symmetry in type II$_A$ string on $K3$, allows to think of phenomena of 
enhancement of symmetry at regions in the moduli space of type II theories 
that are at very strong coupling. In fact, from the analysis in \cite{AGM}, 
topology changing transitions must be taken into account as a present 
mechanism in the singular loci associated to vanishing of the discriminant. 
Some of these transitions are geometrically identical to the process of 
regularization of singularities, again characterized by a Dynkin diagram 
and a set of collapsing $2$-cycles. This comment unifies two apparently 
unrelated facts: topology changing amplitudes, and singularties associated to 
enhanced non abelian symmetries. Even when these phenomena take place in 
very stringy regions of the moduli of type II theory, it is possible to 
use a ``gauge field theory auxiliary model'' to describe them
\cite{KMP}. The interplay between perturbative and non
perturbative enhancement of non abelian gauge symmetries has
been started to be understood on the context of
compactifications of $F$-theory \cite{f} on elliptic fibrations
\cite{ef}, a question also related to the problem of finding
heterotic-heterotic dual pairs.
  
In this notes we will try to attack some of the exciting developments taking 
place around $M$-theory and $F$-theory from a more abstract, but at the 
same time 
simpler, point of view, that arising from the integrable model introduced 
by Donagi and Witten \cite{DW}. The essence of this integrable
model is a two dimensional gauge theory with a Higgs field
living in the adjoint representation defined on a reference
Riemann surface; in what follows we will think of this surface
as the genus one elliptic curve of the $N\!=\!4$ theory
determined by Seiberg and Witten, $E_{\tau}$. The integrable
model, when defined on $E_{\tau}$, can be used to derive the
Seiberg-Witten solution of $N\!=\!2$ supersymmetric gauge
theory. 
  
The relevant part of the integrable model we will extensively
make use of all over this notes is the role played by the $\tau$
moduli on the reference surface: it defines the ``scale'' with
respect to which we ``measure'' the two dimensional gauge
invariant $\mbox{tr}\phi^k$ quantities. 
  
\vspace{2 mm}
  
In these notes evidence will be
presented for interpreting the moduli of Donagi-Witten theory as
the stringy moduli of a Calabi-Yau manifold \cite{GHL}. A reason for such
an interpretation is that in both cases $N\!=\!2$ field theory
is obtained by the same type of blow up procedure. What in the
Calabi-Yau framework is purely stringy, within the Donagi-Witten
approach is interpreted in terms of changes in the moduli of the
``reference'' Riemann surface $E_{\tau}$. The most interesting
part of this construction is that in the integrable model
context, the roles played by $u$ (the value of $\mbox{tr}\phi^2$
relative to the quadratic differential of $E_{\tau}$) and $\tau$
are very symmetric. We can change $u$ or, on the same footing,
modify the scale defined by $\tau$. 
  
It is therefore tempting to consider the duality arising in Donagi-Witten
picture as the same sort of exchange taking place in $F$-theory
elliptic fibrations between the two base \IP$^1$, an exchange at
the root of heterotic-heterotic duality and non perturbative
enhancement of non abelian symmetry. 
  
We feel that if $M$-theory and $F$-theory approaches are deep
and stringy in spirit, the approach based on integrable models
can be complementary replacing the climbing up in dimensions for
the abstract set up of integrability.


\section{Seiberg-Witten Theory.}

For simplicity, and because on this lecture we will only consider the 
case of $SU(2)$, we reduce this brief introduction on Seiberg-Witten 
theory to the simplest case, that of $N\!=\!2$ supersymmetric gauge theory 
with gauge group of rank equal one.
  
The potential for the scalar superpartner $\phi$ is given by
\begin{equation}
V(\phi)=\frac {1}{g^2}\mbox{tr}[\phi,\phi^{\dagger}]^2.
\end{equation}
Vanishing of the potential leads to a flat direction defined by 
\begin{equation}
\phi=\frac {1}{2}a\sigma^3,
\end{equation}
where $a$ is a complex parameter, and $\sigma^3$ is the diagonal Pauli 
matrix. As vacuum states corresponding to values $a$ and $-a$ are 
equivalent, since they are related by the action of the Weyl subgroup, a 
gauge invariant parameterization of the moduli space is defined in terms 
of the expectation value of the Casimir operator,
\begin{equation}
u=<tr \phi^2>=\frac {1}{2}a^2.
\end{equation}
For non vanishing $u$, the low energy effective field theory contains only one 
abelian $U(1)$ gauge field; this gauge field, besides the photon, can 
be described, 
up to higher than two derivative terms, by a holomorphic prepotential 
${\cal F}(a)$. The dual variable is defined through
\begin{equation}
a_D \equiv \frac {\partial {\cal F}}{\partial a}.
\end{equation}

The Seiberg-Witten solution is obtained when an elliptic curve $\Sigma_u$ 
is associated to each value of $u$ in the moduli space; the periods of this 
curve are required to satisfy 
\begin{eqnarray}
  a(u) & = & \oint_A \lambda, \nonumber \\
a_D(u) & = & \oint_B \lambda, 
\label{eq:b4}
\end{eqnarray}
with $A$ and $B$ cycles on the homology basis, and $\lambda$ a meromorphic 
1-form satisfying
\begin{equation}
\frac {d \lambda}{du} =\frac {dx}{y}
\end{equation}
where $\frac {dx}{y}$ is the abelian differential of the curve $\Sigma_u$.
    
From the BPS mass formula for the particle spectrum,
\begin{equation}
M_{BPS} = | \: a n_e + a_D n_m \: |,
\end{equation}
and equation (\ref{eq:b4}), we observe that massless particles will appear 
whenever the homology cycles of $\Sigma_u$ contract to a point.
  
\vspace{2 mm}

Classically, solutions $a(u)$ and $a_D(u)$ are given by the Higgs expressions
\begin{eqnarray}
  a(u) & = & \sqrt{2u}, \nonumber \\
a_D(u) & = & \tau_0 \sqrt{2u},
\end{eqnarray}
with $\tau_0= \frac {\theta}{2\pi}+i \frac {4\pi}{g^2}$ the bare coupling 
constant. The one loop renormalization of the coupling constant leads, in the 
Higgs phase, to
\begin{eqnarray}
  a(u) & = & \sqrt{2u}, \nonumber \\
a_D(u) & = & \frac {2ia}{\pi}\log \left( \frac {a}{\Lambda} \right) + 
\frac {ia}{\pi},
\label{eq:b8}
\end{eqnarray}

Solution (\ref{eq:b8}) is locally correct as a consequence of the 
holomorphy of the prepotential. The global solution, consistent with 
holomorphy, is given by the cycles of the elliptic curve
\begin{equation}
y^2=(x-u)(x-\Lambda^2)(x+\Lambda^2),
\label{eq:swl}
\end{equation}
which leads to the existence of a dual phase where $a(u)$ behaves 
logarithmically, as a consequence of the appearance in the spectrum of a 
charged massless monopole.
  
\vspace{2 mm}
  
The $Sl(2,{\bf Z})$ duality transformations, acting on the vector 
$(a,a_D)$, became a non perturbative symmetry only if they are part of the 
monodromy around the singularities. The monodromy subgroup of $Sl(2,{\bf Z})$ 
for the $SU(2)$ case is given by the group $\Gamma_2$ of unimodular matrices, 
congruent to the identity up to {\bf Z}$_2$.
  
Some points must be stressed, concerning Seiberg-Witten solution:
\begin{itemize}
        \item[{C-1}] Under a soft supersymmetry breaking term, the massless 
        particles appearing at the singularities condense, moving the theory 
        into the confinement (in the case that the condensing particles are 
        monopoles), or the oblique confinement phase (if they are dyons).
        \item[{C-2}] The different phases of the theory, corresponding as 
        mentioned in the above comment to the kind of state giving rise to 
        the singularity, are interchanged by the action of the part of the 
        duality group that is not in the monodromy group,
        \begin{equation}
        Sl(2,{\bf Z})/\Gamma_2.
        \end{equation}
        \item[{C-3}] The elliptic modulus $\tau_u$ of the curve $\Sigma_u$ 
        is given by
        \begin{equation}
        \tau= \frac {d a_D}{da}.
        \end{equation}
        This modulus becomes $0$, $1$ or $\infty$ at the singularities.
        \item[{C-4}] The point $u=0$, corresponding to (classical) 
        enhancement of symmetry, is not a singular point, and therefore no 
        enhancement of symmetry does take place dynamically.
        \item[{C-5}] There exits a global {\bf Z}$_2$ R-symmetry acting on the 
        moduli space by
        \begin{equation}
        u \rightarrow -u.
        \label{eq:b13}
        \end{equation}
        This transformation is part of $Sl(2,{\bf Z})/\Gamma_2$, and it 
        interchanges the monopole and dyon singularities. We can not quotient 
        the moduli space by (\ref{eq:b13}).
        \item[{C-6}] The dynamically generated scale $\Lambda$ of the 
        theory is not 
        a free parameter that can be changed at will. If that was the case, we 
        would be able to consider a family of $SU(2)$ theories with 
        different values of $\Lambda$, with the $\Lambda=0$ theory possessing 
        an enhancement of symmetry singular point. Nevertheless, it is 
        possible 
        to think that once our theory is embedded into string theory, the 
        dilaton can effectively change the value of $\Lambda$, modifying the 
        picture of dynamical enhancement of symmetries. This particular issue 
        will be considered in the rest of this notes.
\end{itemize}


\section{$N\!=\!4$ to $N\!=\!2$ Flow.}
  
Let us start considering the algebraic geometrical description of $N\!=\!4$ 
$SU(2)$ super Yang-Mills \cite{SW2}. For this theory we can define 
a complexified 
coupling constant,
\begin{equation}
\tau=\frac {\theta}{2 \pi} + i \frac { 4 \pi}{g^2}  ,
\end{equation}
on which the duality group $Sl(2,Z)$ acts in the standard way through
\begin{equation}
\tau \rightarrow \frac {a \tau + b}{c \tau + d}.
\end{equation}
As the $\beta$-function for the $N\!=\!4$ theory vanishes, we can take the 
classical solution 
\begin{eqnarray}
    a(u) & = & \sqrt{2 u}, \nonumber \\
a_{D}(u) & = & \tau \sqrt{2 u},
\end{eqnarray}
as exact. The algebraic geometrical description of this solution requires 
finding, for given values of $\tau$ and $u$, an elliptic curve $E_{(\tau,u)}$ 
such that
\begin{eqnarray}
    a(u) & = & \oint_{A} \lambda, \nonumber \\
a_{D}(u) & = & \oint_{B} \lambda,
\end{eqnarray}
where $A$ and $B$ are the two homology cycles, and 
\begin{equation}
\frac {d \lambda}{d u} = \frac {d x}{y} = w,
\end{equation}
with $w$ the abelian differential of the elliptic curve $E_{(\tau,u)}$ (as in 
the $N\!=\!2$ theory of the previous section). This 
problem can be easily solved using the modular properties of 
Weierstrass $\cal P$-function. In a factorized form, the solution is 
given by \cite{SW2}
\begin{equation}
y^2=(x-e_{1}(\tau)u)(x-e_{2}(\tau)u)(x-e_{3}(\tau)u),
\label{eq:swcur}
\end{equation}
in terms of the Weierstrass invariants $e_{i}(\tau)$:
\begin{equation}
        \begin{array}{ccl}
        e_{1}(\tau)= & \frac {1}{3}(\theta_{2}^4(0,\tau)+\theta_{3}(0,\tau))=
        & \frac {2}{3} + 16 q + 16 q^2+ \cdots , \\
        e_{2}(\tau)= & - \frac {1}{3}(\theta_{1}^4(0,\tau)+\theta_{3}^4
        (0,\tau))= & - \frac {1}{3} - 8 q^{1/2} -8 q -32 q^{3/2} - 8q^2+  
        \cdots , \\
        e_{3}(\tau)= & \frac {1}{3} (\theta_{1}^4(0,\tau)-\theta_{2}^4
        (0,\tau))= & - \frac {1}{3}+ 8q^{1/2} - 8q +32 q^{3/2}- 8q^2+ \cdots ,
        \end{array}
\label{eq:winv}
\end{equation}
where we have used Jacobi's $\theta$-functions, and 
$q\equiv e^{2 \pi i \tau}$. By 
an affine transformation, the curve (\ref{eq:swcur}) can be rewritten as 
\begin{equation}
y^2=x(x-1)(x-\lambda(u,\tau)),
\end{equation}
with 
\begin{equation}
\lambda(u,\tau)=\frac {u(e_{3}-e_{1})}{u(e_{2}-e_{1})}.
\label{eq:lambda}
\end{equation}
From equation (\ref{eq:lambda}), we observe that the elliptic modulus of 
the curve 
$E_{(\tau,u)}$ is independent of the value of $u=<tr \phi^2 >$. Interpreting 
$\tau$ as the ``bare'' coupling constant, and the elliptic modulus of 
$E_{(\tau,u)}$ as 
the effective wilsonian coupling constant, the previous fact simply reflects 
the vanishing of the $\beta$-function.
   
Using relations (\ref{eq:winv}), we can now work the weak coupling limit 
$(q \rightarrow 0)$ of (\ref{eq:swcur}):
\begin{equation}
y^2=(x-\frac {2}{3}u)(x+\frac {1}{3}u)^2.
\end{equation}
Redefining $x$ as $(x+ \frac {1}{3}u)$, we get
\begin{equation}
y^2=(x-u)x^2,
\end{equation}
which is precisely the $\Lambda \rightarrow 0$ limit of the Seiberg-Witten 
curve (\ref{eq:swl}).
  
There already exists strong evidence on the Montonen-Olive duality 
invariance \cite{MO} 
of the $N\!=\!4$ gauge theory. Two facts contribute to the existence of this 
symmetry: i) The vanishing of the $\beta$-function, and ii) the correct spin 
content for electrically and magnetically charged BPS-particles. This second 
fact depends on the multiplicities of $N\!=\!4$ vector multiplets\footnote{The 
multiplicities for massless irreducible representations are:

\begin{tabular}{ccl}
        $N\!=\!2$ & (1,2,2) & for spin 1 (vector), \\
                  & (  2,4) & for spin 1/2 (hypermultiplet); \\
        $N\!=\!4$ & (1,4,6) & for spin 1.
\end{tabular}
}. One thing that prevents the existence of exact Montonen-Olive duality 
in $N\!=\!2$ 
is the fact that the $W^{\pm}$'s are in $N\!=\!2$ vector multiplets, 
while the magnetic 
monopoles are in $N\!=\!2$ hypermultiplets. We can easily make the theory 
$N\!=\!4$ invariant by simply adding a massless hypermultiplet in the adjoint 
representation. Naively, we can now have the following physical picture in 
mind: when a soft breaking term is added for the hypermultiplet in the 
adjoint, 
$N\!=\!4$ is broken down to $N\!=\!2$, but still preserving the number of 
degrees 
of freedom which are necessary for having duality, namely the number of 
degrees 
of freedom of the $N\!=\!4$ theory. If the soft breaking mass term goes to 
$\infty$, we obtain at low energy the pure $N\!=\!2$ theory. It looks like if 
in order to have duality for this $N\!=\!2$ theory we should be faced to deal 
with a set of very massive states which have ``a priori'' no interpretation 
in the context of the pure $N\!=\!2$ theory\footnote{At this point we can 
offer an heuristic interpretation in terms of a similar phenomenon. It is well 
known that, in a theory which is anomaly free, when some chiral fermion, that 
contributes to the anomaly cancellation, becomes decoupled some extra state 
appears in the spectrum to maintain the anomaly cancellation right. In our 
context the anomaly condition is replaced by the duality symmetry, and 
anomaly matching conditions by some sort of ``duality'' matching conditions.}. 
  
Motivated by the previous discussion, we will work out in some detail the 
soft breaking $N\!=\!4$ to $N\!=\!2$ flow for the $SU(2)$ gauge theory.
   
The curve describing the massive case was derived in reference \cite{SW2}, and 
is given by 
\begin{equation}
y^2=(x-e_{1}(\tau)\tilde{u}-e_{1}^2(\tau)f)(x-e_{2}(\tau)
\tilde{u}-e_{2}^2(\tau)f)(x-e_{3}(\tau)\tilde{u}-e_{3}^2(\tau)f),
\label{eq:swn4}
\end{equation}
where $f=\frac {1}{4}m^2$, and $\tilde{u}$ is related to $u=<tr \phi^2>$ 
by the 
renormalization
\begin{equation}
\tilde{u}=u- \frac {1}{2}e_{1}(\tau)f.
\end{equation}
The singularities of (\ref{eq:swn4}) are located at 
\begin{equation}
\tilde{u}=e_{1}(\tau)f,e_{2}(\tau)f,e_{3}(\tau)f.
\label{eq:n4loci}
\end{equation}


\subsection{Extended Moduli.}

In order to present the singularities of the massive curve, it is 
convenient to introduce an 
{\em ``extended'' moduli}, parameterized by $\tilde{u}$ and $\tau$. In this 
extended moduli the singularity loci (\ref{eq:n4loci}) have the look 
depicted in Figure 1.


\begin{figure}[htbp]
\centering

        \begin{picture}(400,170)

        \put(100,120){\line(1,0){180}}  \put(140,120){\line(1,-2){50}}
        \put(140,20){\line(0,1){100}}  \put(240,120){\line(-1,-2){50}}

        \put(140,120){\circle*{3}}  \put(190,20){\circle*{3}}

        \put(80,130){$\tau$}  \put(80,115){$0$}  \put(80,25){$i \infty$} 
        \put(200,25){A}
        \put(260,5){$\tilde{u}$}  
        \put(104,50){$e_{1}(\tau)f$} \put(225,50){$e_{2}(\tau)f$}
        \put(170,70){$e_{3}(\tau)f$}
        \put(144,30){I} \put(170,30){II} \put(225,30){III}

        \thicklines
        \put(100,20){\vector(0,1){120}}  \put(80,20){\line(1,0){220}}

        \end{picture}

\caption{{\em The singularity loci (3.14); the point $A$ corresponds 
to $\tilde{u}=-\frac {1}{3}f$ $(u=0)$}.}

\end{figure}
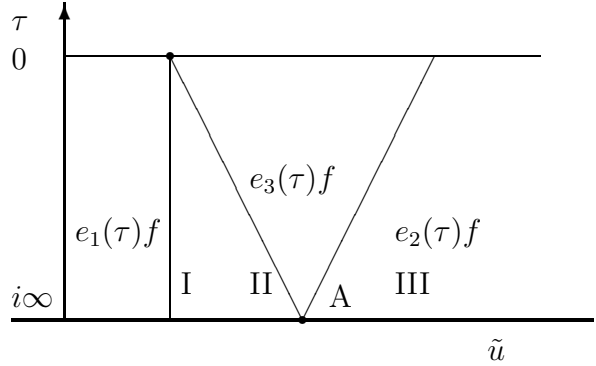

  
The figure is only qualitative, pretending mainly to stress the fact that 
for $\tau=i \infty$ the singularities $e_{2}(\tau)f$ and $e_{3}(\tau)f$ 
coincide, while for $\tau \rightarrow 0$ the coalescing singularities are the 
ones defined by the Weierstrass invariants $e_{1}(\tau)$ and $e_{2}(\tau)$. 
Notice also that in the massless limit $f \rightarrow 0$ the curve 
(\ref{eq:swn4}) 
becomes the $N\!=\!4$ curve (\ref{eq:swcur}). In contrast to what happens for 
the $N\!=\!4$ case, the moduli of the curve (\ref{eq:swn4}) depends on 
$\tau$ and $u$, a fact that reflects the renormalization of the bare coupling 
constant $\tau$. To see the dependence on $u$ and $\tau$ of the moduli of 
(\ref{eq:swn4}) it is convenient to write the curve in the standard form
\begin{equation}
y^2=x(x-1)(x-\lambda(\tilde{u},\tau,f)),
\label{eq:215}
\end{equation}
with
\begin{equation}
\lambda(\tilde{u},\tau,f)=\frac {(e_{3}(\tau)-e_{1}(\tau))(\tilde{u}+
f(e_{3}(\tau)+e_{1}(\tau)))}
        {(e_{2}(\tau)-e_{1}(\tau))(\tilde{u}+f(e_{2}(\tau)+e_{1}(\tau)))}.
\label{eq:ln4}
\end{equation}
It is easy to check that the singularity loci (I,II,III) appearing in Figure 
1 correspond respectively to $\lambda(\tilde{u},\tau,f)=1,\infty,0$.
  
A simple minded approach to Figure 1 is to compare it with a fictitious 
extended 
moduli for the pure $N\!=\!2$ theory, parameterized by $u$ and the 
dynamically generated 
scale $\Lambda$. In this interpretation, the singular loci II and III of 
Figure 1 
will represent the split of the ``classical'' singularity represented by 
the point 
$A$, which corresponds to the point $u=0$ of enhancement of $SU(2)$ symmetry.  
However, this interpretation is too naive, and is not taking into account that 
the pure $N\!=\!2$ theory is only recovered in the $m \rightarrow \infty$ 
limit, 
with $m$ the mass of the extra hypermultiplet in the adjoint representation.
  
\subsection{Extended Moduli and Integrability.}
\label{sec:dw}

The possibility to encode exact information for $N\!=\!2$ systems in terms 
of algebraic geometry opens the way to a connection with integrable
systems \cite{R,MW,DW}. In particular, the extended moduli space 
parameterized by $\tilde{u}$ and $\tau$ can also be 
approached in terms of an integrable system, introduced by 
Donagi and Witten \cite{DW}. 
Let us briefly review this work. 

The integrable model associated with an 
$N\!=\!2$ gauge theory with gauge group $G$ is defined by a bundle 
\begin{equation}
X \rightarrow {\cal U},
\label{eq:dw1}
\end{equation}
with $\dim {\cal U}=r=\mbox{rank} G$, and the fiber $X_{\cal U}$ the jacobian 
$\mbox{Jac} (\Sigma_{\overline{u}})$, where $\Sigma_{\overline{u}}$ is the 
hyper-elliptic curve of genus $g=r$ solving the model in the 
Seiberg-Witten sense \cite{sun} $(\overline{u}=(u_1,\ldots,u_r))$. A Poisson 
bracket structure can be defined on $X$ with respect to which the $u_i$ are 
the set of commuting hamiltonians. Following Hitchin's work \cite{Hi}, we 
can model the integrable system (\ref{eq:dw1}) in terms of a gauge theory with 
gauge group $G$ and a Higgs field $\Phi$ in the adjoint representation of 
$G$, defined on a Riemann surface $\Sigma$. In very synthetic terms, given the 
data $(\Sigma, G,\Phi)$, we define $X$ as the cotangent bundle 
$T^*{\cal M}_{\Sigma}^G$, 
with ${\cal M}_{\Sigma}^G$ the moduli of $G$-connections defined on the 
Riemann surface $\Sigma$. The set of commuting hamiltonians has now a very 
nice 
geometrical interpretation. For $G=SU(n)$, the gauge invariants 
$\mbox{tr}(\phi)^k$ 
define holomorphic $k$-differentials on the Riemann surface $\Sigma$. Now, we 
can expand these gauge invariant Casimirs in terms of a basis of 
$k$-differentials of 
$\Sigma$. The coefficients will define the set of commuting hamiltonians. 
There is, in addition, a simple geometrical way, using the concept of 
ramified coverings (spectral curves), to define on a given Riemann surface 
$\Sigma$ the gauge theory used above to define the integrable model; the 
receipt 
is the following: on a Riemann surface $\Sigma$ we define a holomorphic 
1-differential $\Phi$, that will play the role of the Higgs field, 
valued in the adjoint representation of the gauge group $G$. Given $\Phi$, the 
spectral cover $C$ of $\Sigma $ is then
\begin{equation}
\det (t-\Phi)=0.
\end{equation}
This relation can be written (in the $SU(n)$ case) as
\begin{equation}
t^n+t^{n-2}W_2(\phi) + \cdots + W_n(\phi) =0, 
\label{eq:dw3}
\end{equation}
with $W_k(\phi)$ gauge invariant holomorphic $k$-differentials on $\Sigma$. 
As has already been mentioned, the commuting hamiltonians are given by the 
coefficients 
of $W_k(\phi)$ with respect to the basis of holomorphic $k$-differentials 
of $\Sigma$. Now, the definition on $\Sigma $ of an $SU(n)$ gauge bundle 
$V$ is rather simple: for any point $w \in \Sigma$, from (\ref{eq:dw3}) 
we get $n$ different points $v_i(w)$ in $C$; each of these points is 
associated to a one dimensional eigenspace of $\Phi(w)$ that we will denote 
by $L_{v_i(w)}$. As the point $v_i$ in $C$ is moved, it defines a line 
bundle $L$ on the spectral cover $C$, and a vector bundle $V$ on the 
Riemann surface $\Sigma$:
\begin{equation}
V=\bigoplus_{i=1}^n L_{v_i(w)}.
\end{equation}
However, this vector bundle does not yet define an $SU(n)$ gauge theory on 
$\Sigma$; in order to do so, an extra condition must be imposed: the line 
bundle $\det V$ on $\Sigma $ has to be a trivial bundle. It turns out 
that this 
condition is equivalent to defining the $X_{\overline{u}}$ fiber of the 
corresponding 
integrable system (\ref{eq:dw1}) as the kernel of the map from 
$\mbox{Jac}(C) \rightarrow \mbox{Jac}(\Sigma)$, defined by the map $L 
\rightarrow N(L)$, with $N(L)$ given by 
\begin{equation}
N(L)=\bigotimes_{i=1}^nL_{v_i(w)},
\end{equation}
a line bundle on $\Sigma$ (recall that $L$ was defined as a line bundle on the 
spectral cover $C$).
   
For the $SU(2)$ example, the previous discussion can be materialized in simple 
terms. As reference surface $\Sigma$, we take the $N\!=\!4$ solution 
\begin{equation}
E: \: \:   y^2  =  (x-e_1(\tau))(x-e_2(\tau))(x-e_3(\tau)).
\label{eq:dw6}
\end{equation}
Then, the spectral cover is defined by (\ref{eq:dw6}) and 
\begin{equation}
0 = t^2 -x +\tilde{u}.
\label{eq:dw7}                                         
\end{equation}
This spectral cover is a genus two Riemann surface that we are parameterizing 
by the $\tau$ moduli of (\ref{eq:dw6}), and the $\tilde{u}$ parameter in 
(\ref{eq:dw7}), i.e., by the extended moduli we are considering in this notes. 
The $SU(2)$ curve can now be obtained projecting out in $\mbox{Jac} (C)$ the 
part coming from $\mbox{Jac}(E)$, using the {\bf Z}$_2\times${\bf Z}$_2$ 
automorphisms of the system defined by (\ref{eq:dw6}) and (\ref{eq:dw7}): 
$\alpha: t \rightarrow -t$, $\beta: y \rightarrow -y$.
As the abelian differential of $E$ is given by $\frac {dx}{y}$, the part of 
$\mbox{Jac}(C)$ not coming from $E$ is precisely the $\alpha \beta$ 
{\bf Z}$_2\times${\bf Z}$_2$-invariant part.
  
\vspace{2 mm}
  
After this brief review of Donagi-Witten theory we would like to address 
the attention 
of the reader to a very simple physical interpretation of the role played 
by $\tau$ (the $N\!=\!4$ moduli). What we are going to compare to the 
vacuum expectation values parameterizing the Coulomb phase, the coordinates 
of a point $\overline{u}$, are the coefficients 
defined in the expansion of the $k$-differentials $\mbox{tr} \phi^k$  
(with $\phi$ the Higgs field of the auxiliary gauge theory) in the basis 
of the 
$k$-differentials of the reference Riemann surface $\Sigma$. For $SU(2)$, 
and with $\Sigma=E$, we get a relation of the type 
\begin{equation}
\mbox{tr} \phi^2=u \Phi_{\tau}^{(2)},
\label{eq:dw8}
\end{equation}
with $\Phi_{\tau}^{(2)}$ the quadratic differential on $E$ for the moduli 
value 
$\tau$. {\em The physical way to understand (\ref{eq:dw8}) is that we are 
measuring 
the value of $\mbox{tr}\phi^2$ with respect to the ``scale'' defined in terms  
of the moduli $\tau$ of the $E$ curve\/}\footnote{Even though it was not 
explicitly mentioned, the mass term breaking $N\!=\!4$ to $N\!=\!2$ 
already enters Donagi-Witten construction. In fact, the Higgs field 
$\Phi$ has a pole with residue given by \[ \left( \begin{array}{cccc} 
                                        1 &   &        &  \\
                                          & 1 &        &  \\
                                          &   & \ddots &  \\
                                          &   &        & -(n-1) \end{array} 
                                           \right)m \] 
for $SU(n)$.}. This comment 
will be important later on, and will be at the basis of the use of the 
Donagi-Witten framework to define a stringy representation of the extended 
moduli. In fact, if we think of $\tau$ as defining the scale we use to 
measure $\mbox{tr} \phi^2$, it is natural to think on some relation 
between $\tau$ and the dilaton field.


\subsection{The Soft Supersymmetry Breaking Scale.}

Let us now use $f=\frac {1}{4}m^2$ as the unit scale to measure 
$u=<tr \phi^2>$. In  
order to do so, we define the dimensionless quantity
\begin{equation}
\hat{u} \equiv u/f.
\label{eq:def}
\end{equation}
The first thing to be noticed is that in the pure $N\!=\!2$ decoupling limit, 
$f \rightarrow \infty$, the whole Seiberg-Witten plane $u$ is 
{\em ``blown down''} 
to the point $\hat{u}=0$, i.e., to the point of enhancement of 
symmetry\footnote{Morally speaking, when we measure in units of $f$ and we 
take 
$f \rightarrow \infty $ the theory becomes effectively scale invariant.}. 
Moreover, using again (\ref{eq:def}) we can consider that the point at 
$\infty $ 
of the Seiberg-Witten plane is {\em ``blown up''} into the line $\hat{u}$ 
minus the origin .

Moreover, the scale $\Lambda^2$ is given by $2 q^{1/2} m^2$, and therefore 
in the 
decoupling limit $m \rightarrow \infty$ we can only get a finite scale by 
means of a {\em double scaling limit};
\begin{equation}
\Lambda^2 = \lim_{\begin{array}{c}q \rightarrow 0 \\ m \rightarrow \infty 
\end{array}}
               2 q^{1/2} m^2,
\end{equation}
i.e., in the weak coupling limit $q \rightarrow 0$. Thus, if we work in the 
$(\hat{u},\tau)$-plane, the whole Seiberg-Witten $N\!=\!2$ physics is 
condensed at the singular (enhancement of symmetry) point
$(\hat{u}=0, \tau=i \infty)$. However, it is 
important to stress that we can think of the $(\hat{u},\tau)$-plane without 
imposing any restriction on the value of $f$. In this plane it is only the 
singular point $(\hat{u}=0,\tau=i \infty)$ that admits a pure $N\!=\!2$ 
$SU(2)$ gauge theory interpretation; all other points can only be interpreted 
in terms of the {\em richer} theory containing the hypermultiplet in the 
adjoint 
representation. At this point of the discussion we can face two different 
questions:
\begin{itemize}
        \item[{i)}] How to get the $N\!=\!2$ Seiberg-Witten solution from 
        just the singular point $(\hat{u}=0,\tau = i \infty)$ of the 
        $(\hat{u},\tau)$-plane.
        \item[{ii)}] How to interpret in terms of duality the rest of the 
        $(\hat{u},\tau)$-plane.
\end{itemize}
  
Let us start by considering for the time being, and in very qualitative terms, 
the second of the above interrogants. The action of duality on the 
$(\hat{u},\tau)$-plane 
can be properly defined by the conditions on the moduli
\begin{eqnarray}
S & : & 1-\lambda(\hat{u},\tau)=\lambda(\hat{u}',- \frac {1}{\tau}), 
\nonumber \\
T & : & \frac {1}{\lambda(\hat{u},\tau)}=\lambda (\hat{u}'',\tau+1),
\label{eq:STact}
\end{eqnarray}
defining the $S$ and $T$ action on the $(\hat{u},\tau)$-plane respectively by 
$S:(\hat{u},\tau) \rightarrow (\hat{u}',- \frac {1}{\tau})$, 
$T:(\hat{u},\tau) \rightarrow (\hat{u}'',\tau+1)$. It is easy to verify from 
(\ref{eq:ln4}), (\ref{eq:def}), and the redefinition $\tilde{u}=u-
\frac {1}{2}e_{1}(\tau)f$ that 
\begin{eqnarray}
 \hat{u}' & = & \tau ^2(\hat{u}+ \frac {1}{2}(e_{2}-e_{1})f), \nonumber \\
\hat{u}'' & = & \hat{u}.
\label{eq:utr}
\end{eqnarray}
To derive (\ref{eq:utr}) we have only used the modular properties of the 
Weierstrass invariants $e_{i}(\tau)$, and the fact that
\begin{eqnarray}
1-\lambda(\tilde{u},\tau) & = & \lambda(\tilde{u}^M,- \frac {1}{\tau}), 
\nonumber \\
              \tilde{u}^M & = & \tau^2 \tilde{u},
\end{eqnarray}
that is, $\tilde{u}$ is a modular form of weight two.  
Once we have implemented the action of the duality group $Sl(2,Z)$ on the 
$(\hat{u},\tau)$-plane we observe that the $S$-duality transformations move 
the pure $N\!=\!2$ $SU(2)$ point $A$ into a point with no pure $N\!=\!2$ 
interpretation. This is reminiscent to what happens 
in string theory, where $R-R$ states with no world sheet interpretation are 
necessary to implement duality, and where these states are light only in the 
strong coupling regime, in this case defined by the dilaton.
In summary, and as a temptative answer to the second question 
posed above, we can say that the $(\hat{u},\tau)$-plane beyond the $N\!=\!2$ 
point $(\hat{u}=0,\tau = i \infty)$ is necessary to implement duality. 
But before entering into a more concrete discussion of the previous 
argument, let 
us go first to discuss the first question.


\subsection{The Blow-up Microscope.}

In the previous section we have concentrated the whole $N\!=\!2$ 
Seiberg-Witten 
solution into the singular point $A$ $(\hat{u}=0,\tau = i \infty)$ of the 
$(\hat{u},\tau)$-plane. Now, we face the problem to unravel, just from one 
point, the whole $N\!=\!2$ quantum moduli. By the previous discussion we know 
that the $\hat{u} \neq 0$ has no interpretation in terms of pure $N\!=\!2$ 
$SU(2)$ gauge theory; therefore, to recover Seiberg-Witten quantum moduli we 
will need to add some extra direction parameterized by the pure $N\!=\!2$ 
Seiberg-Witten moduli $u=<tr \phi^2>$. Nicely enough, we have a natural way 
to create this ``extra'' dimension, using simply the fact that the point $A$ 
in the $(\hat{u},\tau)$-plane is a singular point\footnote{Let's recall 
that the parameter $\lambda({\hat u},\tau)$ is ill defined at A.}: this 
singularity can be 
blown-up. To see how this can be done, and get a certain familiarity with the 
blow-up procedure, let us first present some simple examples, referring to 
\cite{H} for a more exhaustive treatment.
   
\vspace{2 mm}   
   
{\bf i) Blowing up a crossing.}
  
\vspace{2 mm}
Let us consider the crossing of the lines $(y=x)$ and 
$(y=0)$. To resolve (blow up) this crossing we define a new variable $v$ 
as $v\equiv y/x$, 
obtaining
\begin{equation}
v=1.
\end{equation}
The coordinate $v$ is a coordinate parameterizing an exceptional divisor $E$ 
which is introduced through the blow up. The result can be represented as in 
Figure 2.


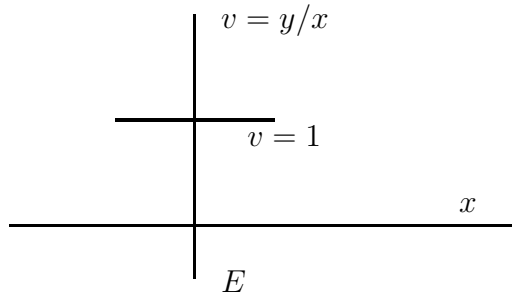
\begin{figure}[htbp]
\centering

     \begin{picture}(400,170)

     \thicklines

     \put(150,20){\line(0,1){100}}  
     \put(80,40){\line(1,0){190}}
     \put(120,80){\line(1,0){60}}

     \put(160,115){$v=y/x$}
     
     \put(250,45){$x$}
     \put(170,70){$v=1$}
     \put(160,15){$E$}

     \end{picture}

\caption{{\em The (single) blow up of the crossing between $y=x$ and $y=0$}.}

\end{figure}

   
\vspace{2 mm}
  
{\bf ii) Blowing up a tangency.}
   
\vspace{2 mm}
We consider now the tangency point $(0,0)$ between the parabola $y=x^2$ and 
the line $y=0$. To blow this tangency up will require a procedure in 
two steps. In the first, the variable $v$ is introduced by the definition
\begin{equation}
v \equiv y/x,
\end{equation}
and we obtain a crossing in the $(v,x)$-plane,
\begin{equation}
v = x.
\end{equation}

To blow up this crossing we repeat the discussion in the above paragraph, 
introducing a new variable $w$,
\begin{equation}
w=v/x,
\end{equation}
so that the original parabola turns into $w=1$. This two stepped process 
has introduced two exceptional divisors $E_{1}$ and 
$E_{2}$, parameterized respectively by $v$ and $w$; the pictorial 
representation is that depicted in Figure 3.


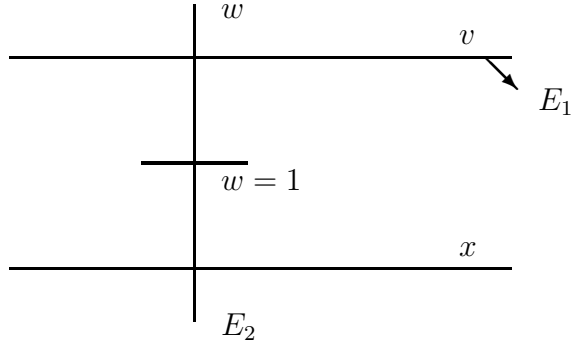
\begin{figure}[htbp]
\centering

     \begin{picture}(400,170)

     \thicklines

     \put(150,20){\line(0,1){120}}  \put(130,80){\line(1,0){40}}
     \put(80,120){\line(1,0){190}}  \put(80,40){\line(1,0){190}}

     \put(260,120){\vector(1,-1){12}}
     
     \put(160,135){$w$} \put(280,100){$E_{1}$}
     \put(250,125){$v$} \put(250,45){$x$}
     \put(160,70){$w=1$} \put(160,15){$E_{2}$}

     \end{picture}

\caption{{\em The double blow up of the tangency point between the 
parabola $y=x^2$ and $y=0$}.}

\end{figure}


\vspace{2 mm}
  
Armed with this artillery we can now blow up the point $A$ in the 
$(\hat{u},\tau)$-plane. 
In the neighborhood of this point, the loci II and III are given by
\begin{eqnarray}
 II & : & \hat{u}(\tau)= 8 q^{1/2}, \nonumber \\
III & : & \hat{u}(\tau)= - 8 q^{1/2}.
\end{eqnarray}
We can now consider, instead of the $\tau$-coordinate, a coordinate $\epsilon$ 
defined by
\begin{equation}
\epsilon \equiv 8 q^{1/2}.
\end{equation}
Using now the blow up of a crossing, as described in i) above, we obtain an 
exceptional divisor $E$ and the picture of Figure 2.


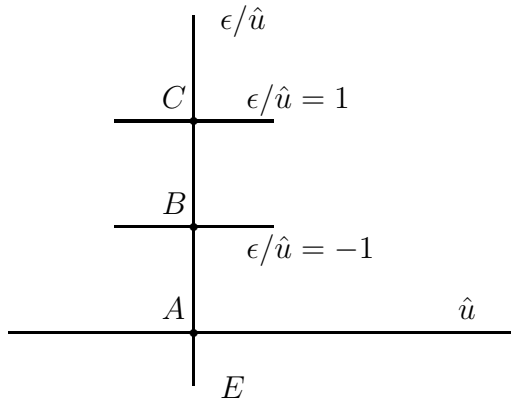
\begin{figure}[htbp]
\centering

     \begin{picture}(400,170)

     \thicklines

     \put(150,20){\line(0,1){140}}  \put(120,80){\line(1,0){60}}
     \put(120,120){\line(1,0){60}}  \put(80,40){\line(1,0){190}}

     \put(150,120){\circle*{3}} \put(150,80){\circle*{3}} \put(150,40)
     {\circle*{3}}

     \put(160,155){$\epsilon/\hat{u}$}
     \put(170,125){$\epsilon/\hat{u}=1$} 
     \put(250,45){$\hat{u}$}
     \put(170,68){$\epsilon/\hat{u}=-1$} 
     \put(160,15){$E$}
     \put(138,45){$A$} \put(138,85){$B$} \put(138,125){$C$}
     \end{picture}

\caption{{\em The double blow up of the tangency point A in the 
$(\hat{u},\tau)$-plane}.}

\end{figure}


Now we can try, following the reasoning at the begining of this section, to 
identify the exceptional divisor $E$ with the Seiberg-Witten quantum moduli 
for $SU(2)$. This is now rather simple using the relation
\begin{equation}
\Lambda^2=8 q^{1/2} f,
\label{eq:lsq}
\end{equation}
with $f=\frac {1}{4}m^2$. In this way, we obtain that the divisor $E$ is 
parameterized by
\begin{equation}
\Lambda^2/u,
\end{equation}
where now $u$ can be identified with the Seiberg-Witten quantum moduli 
parameter, 
with the points $C$ and $B$ in Figure 4 corresponding to the two 
singularities at $u=\pm \Lambda^2$, and the point $A$ to $u=\infty$.
  

\subsection{Double Covering and $R$-Symmetry.}

The action of the $Sl(2,Z)$ modular group on the $(\hat{u},\tau)$-plane was 
defined by equations (\ref{eq:STact}) and (\ref{eq:utr}). Let us now consider 
the effect of the generator $T$,
\begin{equation}
T:(\hat{u},\tau) \rightarrow (\hat{u},\tau+1)
\label{eq:dc1}
\end{equation}
on the coordinate $\epsilon/\hat{u}$ of the exceptional divisor $E$ introduced 
by the blow up; obviously, the action of $T$ induces on $E$ the global 
{\bf Z}$_{2}$ 
transformation
\begin{equation}
\epsilon/\hat{u} \rightarrow - \epsilon/\hat{u},
\label{eq:dc2}
\end{equation}
as $\epsilon=8 q ^{1/2}=8 e^{\pi i \tau}$. 
If we now identify, as was proposed above, the $E$ divisor with the 
Seiberg-Witten 
quantum moduli space for the $N\!=\!2$ $SU(2)$ theory, then the transformation 
(\ref{eq:dc2}) will become the {\bf Z}$_{2}$ global $R$-symmetry 
$u \rightarrow -u $.
  
To quotient by transformation (\ref{eq:dc1}) is equivalent to using 
coordinates 
$(\epsilon^{2},\hat{u})$, instead of $(\epsilon,\hat{u})$. In these new 
coordinates, the loci II and III of Figure 1 can be described, in the 
neighborhood 
of the singular point $(\hat{u}=0,\tau=i \infty)$, by the parabola 
$\hat{u}^{2}=\epsilon^{2}$. 
The tangency can now be blown up through the two stepped process described 
above. 
The two exceptional divisors and coordinates are shown in Figure~5.


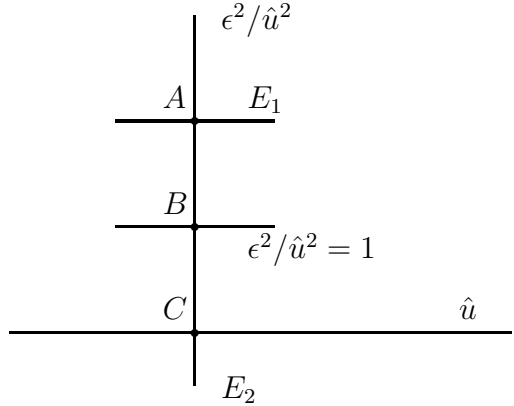
\begin{figure}[htbp]
\centering

     \begin{picture}(400,170)

     \thicklines

     \put(150,20){\line(0,1){140}}  \put(120,80){\line(1,0){60}}
     \put(120,120){\line(1,0){60}}  \put(80,40){\line(1,0){190}}

     \put(150,120){\circle*{3}} \put(150,80){\circle*{3}} \put(150,40)
     {\circle*{3}}

     \put(160,155){$\epsilon^2/\hat{u}^2$}
     \put(170,125){$E_1$} 
     \put(250,45){$\hat{u}$}
     \put(170,68){$\epsilon^2/\hat{u}^2=1$} 
     \put(160,15){$E_2$}
     \put(138,45){$C$} \put(138,85){$B$} \put(138,125){$A$}
     \end{picture}

\caption{{\em Exceptional divisors and coordinates arising from the 
blow up of the tangency point of the parabola $\hat{u}^2=\epsilon^2$ with 
the line $\epsilon=0$}.}

\end{figure}


Using again (\ref{eq:lsq}) we get on $E_{2}$ the dimensionless coordinate 
\begin{equation}
\Lambda^{4}/u^2.
\end{equation}
In Figure 5 the point $B$ corresponds to $\Lambda^4=1$, the point $A$ to 
$u=0$, and the point $C$ to $u=\infty$. The extra divisor at the origin, 
$E_{1}$, appears because we have quotiented by transformation (\ref{eq:dc2}). 
The important point to be stressed here is that the quotient by the global 
{\bf Z}$_{2}$ $R$-symmetry $u \rightarrow -u$ is not allowed in the pure 
$N\!=\!2$ $SU(2)$ gauge theory. Hence, at this level, the quotient by the 
action 
of $T$ is only a formal manipulation. Its real physical meaning will depend on 
our way to understand the $T$ transformation (\ref{eq:dc1}) as a real symmetry 
of the theory.


\subsection{Singular Loci.}

The singular loci, in the $(\hat{u},\tau)$-plane, for the curve 
(\ref{eq:swn4}) 
are given by 
\begin{equation}
        \begin{array}{lcl}
        \hat{{\cal C}}_{\infty}  & \equiv & \{ \tau= i \infty \: / \: 
        \epsilon=0  \}, \\
        \hat{{\cal C}}_{0}       & \equiv & \{ \hat{u}(\tau)=
        \frac {3}{2}e_{1}(\tau) \}, \\
        \hat{{\cal C}}_{C}^{(1)} & \equiv & \{ \hat{u}(\tau)=e_{3}(\tau)
        + \frac {1}{2}e_{1}(\tau) \}, \\
        \hat{{\cal C}}_{C}^{(2)} & \equiv & \{ \hat{u}(\tau)=e_{2}(\tau)
        + \frac {1}{2}e_{1}(\tau) \},  \\
        \hat{{\cal C}}_{1}^+     & \equiv & \{\tau = 0 \: / \: \epsilon=1 
        \}, \\
        \hat{{\cal C}}_{1}^-     & \equiv & \{\tau = 1 \: / \: \epsilon=-1 \},
        \end{array}
\label{eq:loci}
\end{equation}

In the above set, two different types of singular loci can be distinguished: 
$\hat{{\cal C}}_{0}$, $\hat{{\cal C}}_{C}^{(1)}$ and 
$\hat{{\cal C}}_{C}^{(2)}$, describing singularities of the massive curve 
of the $N\!=\!2$ theory with $N\!=\!4$ matter content, and the loci 
$\hat{{\cal C}}_{\infty}$, $\hat{{\cal C}}_{1}^+$ and $\hat{{\cal C}}_{1}^-$, 
related only to the value of $\tau$. 

Some sort of ``duality'' between this two types of loci can already be 
noticed through the $T$-transformation (\ref{eq:dc1}), as (\ref{eq:dc1}) 
interchanges $\hat{{\cal C}}_{1}^+$ with $\hat{{\cal C}}_{1}^-$, and 
$\hat{{\cal C}}_{C}^{(1)}$ with $\hat{{\cal C}}_{C}^{(2)}$. Moreover, 
by (\ref{eq:dw1}) $\hat{{\cal C}}_{0}$ and $\hat{{\cal C}}_{\infty}$ are 
kept fixed.


\section{String Theory Framework.}

\subsection{The Moduli of Complex Structures.}

For Calabi-Yau practitioners the formal similarities between 
the loci (\ref{eq:loci}) and the singular loci of complex structures of 
\IP$_{\{11226\}}$ or of \IP$_{\{11222\}}$ would become immediately 
clear \cite{can,CYY}. We will concentrate our discussion in 
\IP$_{\{11226\}}$. 
  
The mirror of this weighted projective Calabi-Yau space \IP$_{\{11226\}}$ 
is defined by the polynomial\footnote{Strictly, the mirror manifold is
defined by $\{p\!=\!0\}/G$, where $G$ is the group of reparametrization
symmetries of $p$ \cite{can}.}
\begin{equation}
p=z_{1}^{12}+z_{2}^{12}+z_{3}^{6}+z_{4}^{6}+z_{5}^{2}-12 \psi 
z_{1}z_{2}z_{3}z_{4}z_{5}- 2 \phi z_{1}^{6}z_{2}^{6}.
\label{eq:11226}
\end{equation}
The moduli of complex structures, parameterized by $\psi$ and $\phi$,
presents a {\bf Z}$_{12}$ global symmetry $A$:
\begin{eqnarray}
\psi \rightarrow \alpha \psi, \nonumber \\
\phi \rightarrow - \phi,
\label{eq:z12}
\end{eqnarray}
where $\alpha$ is such that $\alpha^{12}=1$. The singular loci in the 
compactified moduli space are given by 
\begin{equation}
        \begin{array}{lcl}
        {\cal C}_{con}    & \equiv & \{ 864 \psi^6+ \phi = \pm 1 \}, \\
        {\cal C}_{1}      & \equiv & \{ \phi = \pm 1 \}, \\
        {\cal C}_{\infty} & \equiv & \{ \phi, \psi = \infty \}, \\
        {\cal C}_{0}      & \equiv & \{ \psi = 0 \}.
        \end{array}
\label{eq:lc2}
\end{equation}
The locus ${\cal C}_{0}$ appears as the result of quotienting by the 
{\bf Z}$_{12}$ symmetry (\ref{eq:z12}). In fact, $g^2$ fixes the whole 
line $\psi=0$. Defining the $A$-invariant quantities 
\begin{equation}
\xi \equiv \psi^{12}, \: \: \: \: \: \: \eta \equiv \psi^6 \phi, 
\: \: \: \: \: \: \zeta \equiv \phi^2,
\end{equation}
the moduli space can be defined as the affine cone in {\bf C}$^3$ given by 
the relation 
\begin{equation}
\xi \zeta = \eta^2.
\end{equation}

To compactify this space we can embed {\bf C}$^3$ into \IP$^3$, with 
homogeneous 
coordinates $[\hat{\xi},\hat{\eta},\hat{\zeta},\hat{\tau}]$, in such a way 
that $\xi=\hat{\xi}/\hat{\tau}$, $\eta=\hat{\eta}/\hat{\tau}$, 
$\zeta=\hat{\zeta}/\hat{\tau}$. 
Now, the compactified moduli is defined by the projective cone ${\cal Q}$, 
defined as 
\begin{equation}
\hat{\xi} \hat{\zeta} = \hat{\eta}^2, 
\end{equation}
and the loci (\ref{eq:lc2}) become
\begin{equation}
        \begin{array}{lcl}
        {\cal C}_{con}    & \equiv & {\cal Q} \bigcap \{ (864)^2 \hat{\xi}+ 
        \hat{\zeta} + 1728 \hat{\eta} - \hat{\tau}= 0 \}, \\
        {\cal C}_{1}      & \equiv & {\cal Q} \bigcap \{ \hat{\eta} - 
        \hat{\tau} = 0 \}, \\
        {\cal C}_{\infty} & \equiv & {\cal Q} \bigcap \{ \hat{\tau} = 0 \}, \\
        {\cal C}_{0}      & \equiv & {\cal Q} \bigcap \{ \hat{\xi}=\hat{\eta} 
        = 0 \}.
        \end{array}
\label{eq:lccone}
\end{equation}
  
Notice for instance that in the compactified moduli the loci ${\cal C}_{1}$ 
and ${\cal C}_{\infty}$ meet at the point $[\hat{\xi}=1,
\hat{\eta}=0,\hat{\zeta}=0,\hat{\tau}=0]$ of \IP$^3$. 
  
The toric representation (see Appendix) of the projective cone ${\cal Q}$ is 
given in Figure~6.


\begin{figure}[htbp]
\centering

        \begin{picture}(400,170)

        \thicklines

        \put(150,60){\vector(1,0){60}} \put(150,60){\vector(0,1){60}}     
        \put(150,60){\vector(-2,-1){70}}

        \put(200,40){$(1,0)$} \put(158,112){$(0,1)$} 
        \put(70,10){$(-2,-1)$}

        \put(202,100){I} \put(160,20){II} \put(90,100){III}

        \end{picture}

\caption{{\em Toric diagram of the projective cone ${\cal Q}$, containing 
the three different coordinate patches}.}

\end{figure}
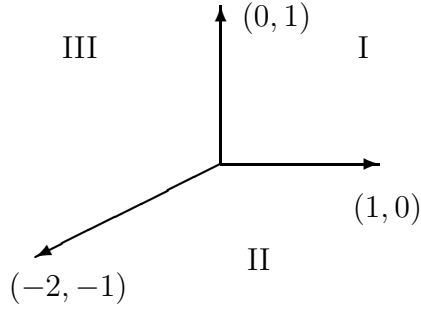


For the coordinates in chart I we introduce the variables 
\begin{equation}
\frac {1}{x} =- \frac {864 \psi^6}{\phi}, \: \: \: \: \: \: \: \: 
y = \frac {1}{\phi^2}.
\label{eq:141}
\end{equation}
  
In this chart the four singularity loci are those depicted in Figure~7.


\begin{figure}[htbp]
\centering

        \begin{picture}(400,170)

        \put(100,120){\line(1,0){180}}  \put(140,120){\line(1,-2){49}}
        \put(140,10){\line(0,1){120}}  \put(242,120){\line(-1,-2){49}}
        \put(191,21){\oval(4,2)[b]}

        \put(191,20){\circle*{3}} \put(140,120){\circle*{3}}

        \put(100,130){$y$}  \put(130,5){$0$} 
        \put(190,5){1} 
        \put(260,5){$1/x$}  
        \put(104,50){${\cal C}_0$} \put(225,50){${\cal C}_{con}$}
        
        \put(240,20){\vector(1,1){20}} \put(270,45){${\cal C}_{\infty}$}
        \put(240,120){\vector(1,1){20}} \put(270,145){$y=1$}

        \thicklines

        \put(80,20){\line(1,0){220}}

        \end{picture}

\caption{{\em Singular loci of the compactified moduli space as seen in 
coordinate chart I}.}

\end{figure}
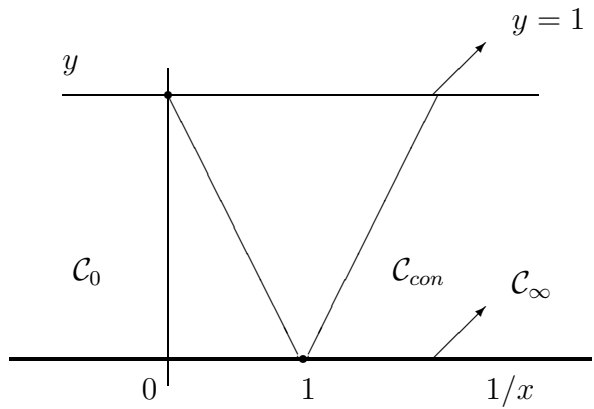


When passing to the coordinate chart II, parametrized by coordinates 
$(x,x^2y)$, the four singularity loci will appear in a different fashion, 
as shown in Figure~8.


\begin{figure}[htbp]
\centering

        \begin{picture}(400,170)

        \put(140,120){\line(1,-2){49}}
        \put(140,10){\line(0,1){120}}  \put(242,120){\line(-1,-2){49}}
        \put(191,21){\oval(4,2)[b]}
        \put(140,21){\oval(4,2)[b]} \put(89,120){\line(1,-2){49}}
        \put(191,120){\line(-1,-2){49}}
        \put(240,20){\line(0,1){40}} 

        \put(191,20){\circle*{3}} \put(140,20){\circle*{3}}

        \put(150,130){$x^2y$}  \put(138,5){$0$} 
        \put(190,5){1} 
        \put(260,5){$x$}  
        \put(245,50){${\cal C}_0$} \put(188,50){${\cal C}_{con}$}
        \put(80,50){${\cal C}_1$}
        \put(238,5){$\infty$} 
        \put(270,20){\vector(1,1){20}} \put(300,45){${\cal C}_{\infty}$}

        \thicklines

        \put(80,20){\line(1,0){220}}

        \end{picture}

\caption{{\em Singular loci of the compactified moduli space as seen in 
coordinate chart II}.}

\end{figure}
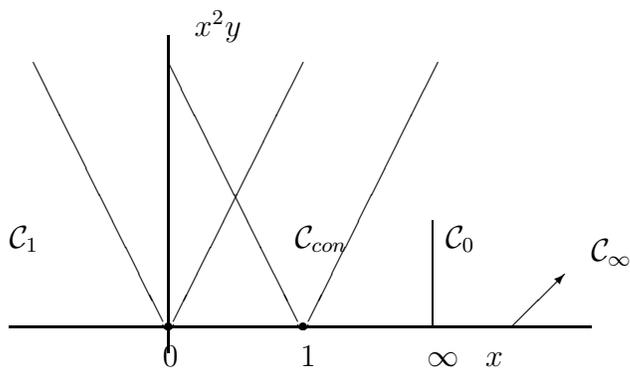


One should notice that in this chart the locus ${\cal C}_{1}$ is 
tangent, at the origin, to the locus ${\cal C}_{\infty}$. This tangency can 
therefore 
be blown up in the usual way (see Figure 9).


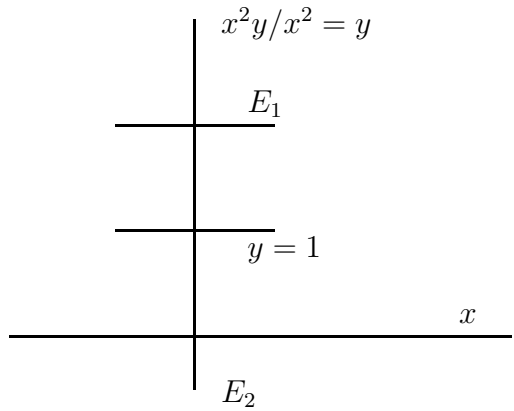
\begin{figure}[htbp]
\centering

        \begin{picture}(400,170)

        \thicklines

     \put(150,20){\line(0,1){140}}  \put(120,80){\line(1,0){60}}
     \put(120,120){\line(1,0){60}}  \put(80,40){\line(1,0){190}}

     \put(160,155){$x^2y/x^2=y$}
     \put(170,125){$E_1$} 
     \put(250,45){$x$}
     \put(170,70){$y=1$} 
     \put(160,15){$E_2$}

        \end{picture}

\caption{{\em Blow up of the tangency point of ${\cal C}_{1}$ to 
${\cal C}_{\infty}$}.}

\end{figure}


It is clarifying to point out that the blow up of Figure~9 can be 
constructed in a toric way, giving rise to 
the toric vectors shown in Figure~10.


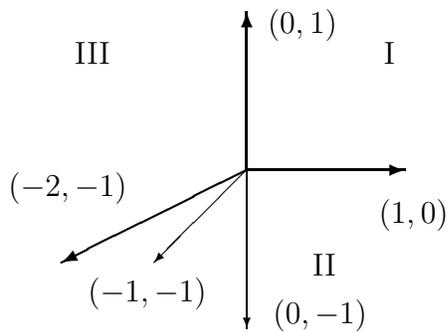
\begin{figure}[htbp]
\centering

        \begin{picture}(400,170)
        
        \put(150,60){\vector(0,-1){60}} \put(150,60){\vector(-1,-1){35}}

        \thicklines

        \put(150,60){\vector(1,0){60}} \put(150,60){\vector(0,1){60}}     
        \put(150,60){\vector(-2,-1){70}} 

        \put(200,40){$(1,0)$} \put(158,112){$(0,1)$} 
        \put(60,50){$(-2,-1)$} \put(90,10){$(-1,-1)$} 
        \put(160,2){$(0,-1)$}

        \put(202,100){I} \put(175,20){II} \put(85,100){III}

        \end{picture}

\caption{{\em Toric diagram of the blow up shown in Figure~9. The divisors 
$E_{1}$ and $E_{2}$ correspond, respectively, to the toric diagram 
vectors $(-1,-1)$ and $(0,-1)$}.}

\end{figure}


\subsection{Back to the $(\hat{u},\tau)$-plane.}
\label{sec:27}

Inspired by the previous discussion, let us come back to the 
$(\hat{u},\tau)$-plane. 
We will introduce a new variable $\tilde{u}$ defined by 
\begin{equation}
\tilde{u} \equiv \hat{u}/\epsilon.
\label{eq:49}
\end{equation}
Now, let us quotient by the global transformation 
\begin{eqnarray}
\tilde{u} & \rightarrow & - \tilde{u}, \nonumber \\
\epsilon  & \rightarrow & - \epsilon,
\label{eq:dc14}
\end{eqnarray}
which is nothing but the $T$ transformation defined 
in (\ref{eq:dc1}). Following the same steps as for the Calabi-Yau moduli 
space, we introduce the variables
\begin{equation}
\xi \equiv \tilde{u}^2,\: \: \: \: \: \: \eta \equiv 
\frac {\tilde{u}}{\epsilon},
\: \: \: \: \: \: \zeta \equiv \frac{1}{\epsilon^2},
\label{eq:dc15}
\end{equation}
satisfying again just the same sort of relation,
\begin{equation}
\xi \zeta = \eta^2.
\label{eq:dc16}
\end{equation}
Now, we embed the cone (\ref{eq:dc16}) into \IP$^3$, using homogeneous 
coordinates 
$[\hat{\xi},\hat{\eta},\hat{\zeta},\hat{\tau}]$. It is now obvious that after 
this compactification the loci ${\cal C}_{\infty}=\{ \tau=i \infty \}$ and 
${\cal C}_{1}=\{\tau = 0\}$ become 
\begin{equation}
        \begin{array}{lcl}
        {\cal C}_{1}      & \equiv & {\cal Q} \bigcap \{ \hat{\eta} - 
        \hat{\tau} = 0 \}, \\
        {\cal C}_{\infty} & \equiv & {\cal Q} \bigcap \{ \hat{\tau} = 0 \},
        \end{array}
\label{eq:dc17}
\end{equation}
with ${\cal Q}$ the projective cone defined by $\hat{\xi} 
\hat{\zeta}=\hat{\eta}^2$. 
In this approach, coordinates in chart I are $\frac {1}{\zeta}=\epsilon^2$, 
and $\frac {\eta}{\zeta}=\hat{u}$. In chart II, the coordinates will be 
$(\frac {\zeta}{\eta}=\frac {1}{\hat{u}},\frac {\zeta}{\eta^2}=
\frac {\epsilon^2}
{\hat{u}^2})$; obviously, in this chart we notice that ${\cal C}_{1}=
\{\epsilon^2=1\}$ is tangent to ${\cal C}_{\infty}$ at the origin, i.e., 
for $\{\hat{u}=\infty\}$. If we blow this tangency up we get two 
exceptional divisor $E_1$ and $E_2$ parameterized respectively by  
$\epsilon^2$ and $\epsilon^2/\hat{u}$.
  
Notice now that the extended moduli space we started with, namely the 
$(\hat{u},\epsilon)$-plane, is a double covering of the chart I of the 
compactified moduli space defined by the projective cone ${\cal Q}:
\hat{\xi} \hat{\zeta}= \hat{\eta}^2$ for $\xi$, $\eta$ and $\zeta$ as 
given in equation (\ref{eq:dc15}). 

\begin{center}
\begin{tabular}{|ccc|}      \hline
$(\hat{u},\epsilon)$-plane & $\longrightarrow$ & $(\hat{u},\epsilon^2)$ = 
chart I of ${\cal Q}$ $[\hat{\xi} \hat{\zeta}= \hat{\eta}^2 ]$ \\ 
                  & {\bf double}      & $\uparrow$ \\
                  & {\bf cover}       & Compactified moduli quotiented by: \\
                  &                   & $T:(\hat{u},\tau) \rightarrow 
(\hat{u},\tau+1)$ \\ \hline
\end{tabular}
\end{center}

If we now consider the double cover of chart II, as defined by coordinates 
$(1/\hat{u},\epsilon/\hat{u})$, we will get a crossing between 
${\cal C}_1^{\pm}$ and ${\cal C}_{\infty}$. By blowing this crossing up, 
we get only 
one exceptional divisor $E$, this time parameterized by $\epsilon$. In the 
$(\hat{u},\epsilon)$-plane, this exceptional divisor can be identified 
with the line $\hat{u}=\infty$.

At this point, we should come back to our discussion in section \ref{sec:dw}. 
The Donagi-Witten integrable model framework is a natural set up for 
interpreting the $\tilde{u}$ variable introduced in (\ref{eq:49}). In fact, 
when we change the value of the $\tau$ moduli of $E$ in equation 
(\ref{eq:dw8}), we 
are effectively changing the value of $\mbox{tr}\phi^2$ as it is measured 
in the 
new unit obtained by varying the $\tau$ moduli; this is the same as the 
phenomena we have in equation (\ref{eq:49}) when we change the value of 
$\epsilon$ with fixed value of $\hat{u}$. In fact, in the 
$(\tilde{u},\epsilon)$ 
variables used in the compactification of the moduli, the lines 
of fixed $\hat{u}$ in the $(\hat{u},\epsilon)$-plane are parameterized by 
$\tilde{u}$. Using $\tilde{u}$ as the dimensionless parameter, we have 
two candidates for Seiberg-Witten moduli: the divisor obtained from the 
blow up of the tangency point at $\hat{u}=0$, $\tau=i \infty$, and any line 
of fixed $\hat{u}$, in which $\tilde{u}$ changes depending on the value of 
$\tau$ $(\epsilon)$; are both lines candidates to $SU(2)$ theory, and, if so, 
what is the (different) physics they are representing? In the last part of 
these 
notes, we will argue that both lines describe an $SU(2)$ theory. However, 
before that discussion we will proceed to compare the two moduli spaces 
more carefully.


\subsection{Comparing Blow Ups.}

Let us now compare the blow up in Figure 5 and the blow up of the 
tangency point 
$(x=1,y=0)$ between ${\cal C}_{con}$ and ${\cal C}_{\infty}$. In the 
second case 
the parabola is defined by 
\begin{equation}
y= \left( \frac {1-x}{x} \right) ^2,
\end{equation}
and the blow up is that in Figure 11,


\begin{figure}[htbp]
\centering

        \begin{picture}(400,170)

     \thicklines

     \put(150,20){\line(0,1){140}}  \put(120,80){\line(1,0){60}}
     \put(120,120){\line(1,0){60}}  \put(80,40){\line(1,0){190}}
     
     \put(160,155){$yx^2/(1-x)^2$}
     \put(170,125){$yx/(1-x)$} 
     \put(250,45){$(1-x)/x$}
     \put(170,68){$yx^2/(1-x)^2=1$} 
     
        \end{picture}

\caption{{\em Blow up of the tangency point of ${\cal C}_{con}$ with 
${\cal C}_{\infty}$}.}

\end{figure}
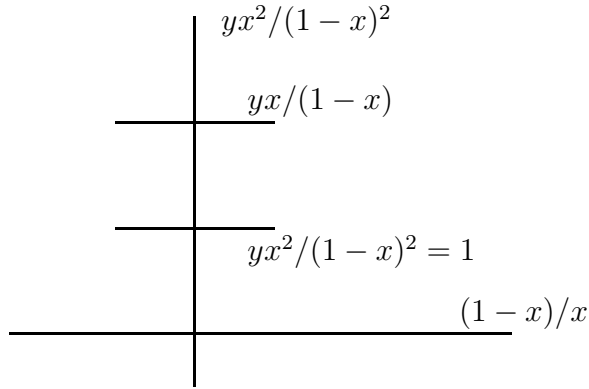


Identifying the two blow ups, we get the following relation:
\begin{equation}
\frac {yx^2}{(1-x)^2}=\frac {\epsilon^2}{\hat{u}^2}.
\label{eq:dc18}
\end{equation}
From (\ref{eq:dc18}), we get the identifications 
\begin{eqnarray}
      y & = & \epsilon^2 \\             
\hat{u} & = & \pm \frac {1-x}{x}.
\label{eq:149}
\end{eqnarray}

As it can be seen from equation (\ref{eq:dc18}) the blow up of the conifold 
singularity produces an exceptional divisor parameterized by 
$1/{\tilde{u}^2}$, 
which can only be related to the Seiberg-Witten moduli plane if we 
first perform the quotient by the {\bf Z}$_2$ $R$-symmetry
\begin{equation}
\tilde{u} \rightarrow -\tilde{u}.
\label{eq:z1}
\end{equation}
When we considered the same problem from the point of view of the 
$(\hat{u},\epsilon)$-plane, 
we notice that transformation (\ref{eq:z1}) was induced by the $T$ action 
$(\hat{u},\tau) \rightarrow (\hat{u},\tau +1)$, which does not modify the 
vacuum expectation value $\hat{u}$. We then conclude that the $T$ 
transformation 
$(\hat{u},\epsilon) \rightarrow (\hat{u},-\epsilon)$, having nothing to do 
with a {\bf Z}$_2$ $R$-symmetry, {\em induces}, on the $\tilde{u}$ variable 
obtained from the blow up, a {\bf Z}$_2$ $R$-symmetry (which is the global 
transformation remaining after taking into account instantonic effects on 
the $SU(2)$ theory). In a simple scheme,

\begin{center}
\begin{tabular}{|ccccc|}      \hline
      & $(\hat{u},\tau)$   & $\longrightarrow$    & $\tilde{u}$     &     \\
 $T:$ & $\downarrow$       & {\bf blow up}        & $\downarrow$    &  
{\bf Z}$_2$ $R$-symmetry \\
      & $(\hat{u},\tau+1)$ & {\bf variable}       & $- \tilde{u}$   &     \\
\hline
\end{tabular}
\end{center}

It must also be noticed that the blow up in chart II of the tangency between 
${\cal C}_1$ and ${\cal C}_{\infty}$ introduces the exceptional divisor 
$E_2$ (see Figure~9), parameterized by $y$, i.e., $E_2=\{x=0\}$. From 
(\ref{eq:149}) we see that $x=0$ corresponds to $\hat{u}=\infty$, in agreement 
with our discussion in section \ref{sec:27}, concerning the tangency between 
$\hat{{\cal C}_1}$ and $\hat{{\cal C}_{\infty}}$ in the 
$(\hat{u},\epsilon)$-plane 
compactified by quotienting by (\ref{eq:dc14}).
In order to get some insight on the physical meaning of these identifications, 
we will first need to comment on the concept of heterotic-type II dual pairs.


\subsection{Heterotic-Type II Dual Pairs.}

In what follows, a very succinct definition of heterotic-type II dual pairs 
will be presented, concentrating in the part of the string 
moduli space associated to vector excitations.
   
A heterotic-type II dual pair is given by a pair $(X,Y)$ of Calabi-Yau 
manifolds, such that the quantum moduli of an heterotic string compactified 
on $X$ coincides with the classical moduli of K\"ahler structures of $Y$. 
The necessary condition for having a dual pair is 
\begin{equation}
b_{1,1}(Y)=b_{2,1}(Y^*)=V,
\label{eq:dc20}
\end{equation}
where $b_{1,1}(Y)$ and $b_{2,1}(Y^*)$ denote Hodge numbers of the manifold 
$Y$ and its mirror\footnote 
{Two Calabi-Yau spaces $Y$ and $Y^*$ are said to constitute a mirror pair if 
they correspond to the same conformal field theory, i.e., when taken as 
target spaces for two dimensional nonlinear $\sigma$-models they give rise 
to isomorphic $N\!=\!2$ superconformal field theories; this isomorphism maps 
the complex structure moduli of $Y$ to the K\"{a}hler moduli of $Y^*$, and 
vice versa. A necessary condition for $Y$ and $Y^*$ to be a mirror pair 
is that their Hodge numbers satisfy $b_{p,q}^{Y}=b_{d-p,q}^{Y^*}$.}
$Y^*$ respectively, and where $V$ stands for 
the number of massless vectors for the heterotic string compactified on 
$X$\footnote{An analogous condition to (\ref{eq:dc20}) should be satisfied
for the part of the moduli space coming from massless neutral
hypermultiplets .}. 
We are not taking into account in (\ref{eq:dc20}) the graviphoton as an extra 
massless vector, while the vector superpartner of the dilaton its included.
  
In more precise terms, the idea of heterotic-type II dual pairs can be 
introduced 
using prepotentials. Let us denote by ${\cal F}^{het}(T_{i},S)$ the 
prepotential 
at tree level of the heterotic string compactified on the manifold $X$, with 
$T_{i}$, $i=1, \ldots, 4$, representing the scalar component of the massless 
vector fields, and $S$ the dilaton. In the same way as for the rigid case, 
presented 
in section 2, this prepotential will become quantum mechanically corrected 
by one loop perturbative effects, and by non perturbative corrections:
\begin{equation}
{\cal F}^{het}(T_{i},S)={\cal F}_{0}^{het}(T_{i},S)+{\cal
F}_{one \: \: loop}^{het}(T_i)
+{\cal F}_{non-perturbative}^{het}(T_i,S).
\end{equation}

Let us denote by ${\cal F}^{II}(t_j)$, $(j=1, \ldots, n+1)$, the prepotential 
characterizing the special K\"{a}hler geometry of the moduli space of 
complex structures 
of some Calabi-Yau space $Y$. The pair $(X,Y)$ of Calabi-Yau manifolds will 
define a heterotic-type II dual pair if there exits a map 
$(T_i,S) \rightarrow (t_j)$ such that ${\cal F}^{II}(t_i)=
{\cal F}^{het}(T_i,S)$.
  
The main difficulty in finding heterotic-type II dual pairs or, in other 
words, 
finding the heterotic dual version of a type II string compactified on some 
Calabi-Yau manifold $Y$, is to discover which variable $t_j$ corresponds to 
the heterotic dilaton $S$. A partial solution to this question comes from 
requiring $Y$ to be a $K_3$-fibration.
  
Let us assume we have a type II$_A$ string compactified on $Y$ (type II$_B$ 
on its mirror image $Y^*$). Then, in the large radius limit we can 
characterize 
the corresponding point in the moduli ${\cal M}$ of K\"{a}hler structures 
($\dim {\cal M}=b_{1,1}(Y)=b_{2,1}(Y^*)$) by a K\"{a}hler form in $H_2(Y)$:
\begin{equation}
B+iJ= \sum (B+iJ)_{j}e_j \equiv \sum t_j e_j.
\end{equation}
(On this paragraph we are following presentation in refence
\cite{AL}). To each element $e_j$ in $H_2(Y)$ we can associate an element 
in the dual 
$H_4(Y)$, i.e., a divisor $D_j$ of complex codimension one. 
The holomorphic prepotential for 
the type II$_A$ theory is then given by
\begin{equation}
{\cal F}^{II}=- \frac {i}{6} \sum_{j_1,j_2,j_3} (D_{j_1} \wedge D_{j_2} \wedge 
D_{j_3}) t_{j_1}t_{j_2}t_{j_3} + \cdots,
\label{eq:dc22}
\end{equation}
where the dots stand for instanton corrections, and $(D_{j_1} \wedge D_{j_2} 
\wedge
D_{j_3})$ is the intersection product, as homology classes, of these divisors. 
Now, we can try to discover which particular divisor $D_S$ corresponds 
to the heterotic dilaton. From Peccei-Quinn symmetry we can fix the tree 
level part of the heterotic prepotential,
\begin{equation}
{\cal F}_0^{het}(T_i,S)= \sum S C_{jk} T_j T_k.
\label{eq:dc23}
\end{equation}
Comparing (\ref{eq:dc22}) and (\ref{eq:dc23}) we get the following constraints 
on the ``dilaton'' divisor $D_S$:
\begin{eqnarray}
D_S \wedge D_S \wedge D_S & = & 0, \nonumber \\
D_S \wedge D_S \wedge D_j & = & 0, \forall j. 
\label{eq:23}
\end{eqnarray}
From the second of the above constraints, we notice that
\begin{equation}
D_S \wedge D_S = 0.
\end{equation}
Two more extra constraints can be derived for $D_S$:
\begin{equation}
e_S \wedge C \geq 0,
\label{eq:dc26}
\end{equation}
for any algebraic curve $C$, and 
\begin{equation}
D_S . c_2(X) =24.
\label{eq:dc27}
\end{equation}
Condition (\ref{eq:dc26}) implies that if the K\"{a}hler form $J$ is in the 
K\"{a}hler cone, then $J'=J+\lambda e_S$ is also in the K\"{a}hler cone. 
Condition (\ref{eq:dc27}) is more involved, and takes into account higher 
derivative terms.
  
The previous set of conditions on $D_S$ implies that $Y$ should be a 
$K_3$-fibration 
with base space \IP$^1$, and fiber a $K_3$-surface. In particular, it also 
means 
that the heterotic dilaton has a geometrical interpretation as the size of 
the fiber \IP$^1$ \footnote{More precisely, from conditions (\ref{eq:23}), 
(\ref{eq:dc26}) and (\ref{eq:dc27}) it is possible to conclude that 
$D_S$, as an element in $H_4$, is given by the $K_3$-surface itself, and 
thus the corresponding cohomology is given by the size of the basis \IP$^1$.}.
Once we have discovered the coordinate $t_j$ for the 
dilaton, we can use the inverse mirror map to define the relation between 
the heterotic dilaton and the complex structure moduli of the mirror manifold.

Let us consider in particular the example proposed by Kachru and
Vafa \cite{KV}. The Calabi-Yau 
space is the quintic in \IP$_{\{11226\}}$, and its mirror $Y^*$ is the one 
defined by the polynomial (\ref{eq:11226}). Using the results in Appendix 
\ref{sec:msing} 
(see equation (\ref{eq:p19})), we get the relation
\begin{equation}
S= \frac {1}{2\pi i} \log \left( \frac {2-z-2\sqrt{1-z}}{z} \right),
\label{eq:160}
\end{equation}
with $z$ defined in (\ref{eq:p16}),
\begin{equation}
z=\frac {1}{\phi^2}.
\end{equation}
Expanding (\ref{eq:160}) around $z=0$, i. e., in the weak coupling limit 
$\phi \rightarrow \infty$, we get
\begin{equation}
S=\frac {1}{2\pi i} \log \left( \frac {z}{4} \right)
\end{equation}
which, in terms of the variable $y$ defined in (\ref{eq:141}), becomes the 
Kachru-Vafa relation \cite{KV}
\begin{equation}
y \simeq e^{- 2\pi i S}.
\end{equation}
Combining now this equation with that from (\ref{eq:149}), obtained by 
comparing the blow up around the conifold point, we get the stringy 
interpretation of the $N\!=\!4$ parameter $\tau$ in terms of the 
dual heterotic dilaton.
  
This identification sends some light on our previous discussion on 
Donagi-Witten theory. In fact, what we identify with the heterotic dilaton is 
the moduli of the reference curve $E_{\tau}$, in terms of which we were 
``measuring'' the two dimensional Higgs gauge invariant quantity 
$\mbox{tr}\phi^2$ (see equation (\ref{eq:dw8})).

Once we have identified the moduli corresponding to the heterotic dilaton, 
we can, in the two moduli case, identify the heterotic $T$ with the second 
generator of $H_2(Y)$. Using again the inverse mirror map we get, for 
$Y=\IP_{\{11226\}}$,
\begin{equation}
x= \frac {1728}{j(T)} + \cdots 
\label{eq:dc30}
\end{equation}
in the large $S$ limit. From (\ref{eq:dc30}), we can identify the point 
$(x=1,y=0)$, where ${\cal C}_{con}$ is tangent to ${\cal C}_{\infty}$, as 
the corresponding to the enhancement of symmetry at $T=i$: 
$U(1) \rightarrow SU(2)$ for the heterotic 
string compactified on $T^2\times  K_3$. In our comparison between the 
$(x,y)$-plane and the $(\hat{u},\tau)$-plane of the $N\!=\!2$ theory with 
$N\!=\!4$ matter content, this point of enhancement of symmetry 
corresponds to $(\tau=i \infty,\hat{u}=0)$.   
Moreover, the quantum moduli space for $N\!=\!2$ $SU(2)$ Yang-Mills theory
can be exactly recovered in the point particle limit of the string
by blowing up the tangency point $(x=1,y=0)$ (see Figure 11)
and identifying the second divisor introduced by the blow up $E_2$
with the Seiberg-Witten moduli space according to (4.15) \cite{KKLMV}.


\subsection{Non Perturbative Enhancement of Gauge Symmetry.}

A natural question we can immediately make ourselves is whether there exists 
any other point in the $(x,y)$-plane, not necessarily in the perturbative 
region $y \rightarrow 0$, which can be interpreted in terms of an 
enhancement of symmetry. Following the general philosophy we have learned 
in the Seiberg-Witten analysis of the rigid theory, we should expect this 
thing 
to take place whenever an {\em electrically\/} charged particle becomes 
massless. 
In the case of the special geometry of the $(x,y)$ moduli, there are two 
special coordinates $t_1$ and $t_2$. They are associated with the two 
$U(1)$ vector fields corresponding to $b_{1,1}(\IP_{\{11226\}})=2$. Moreover, 
from the BPS mass formula, we can get massless electrically charged 
particles with respect 
to each of these $U(1)$'s at points where some of these variables vanish. 
In particular, for $(x=0,y=1)$ we have, from (\ref{eq:dc30}), that 
$t_2=\frac {T-i}{T+i}$ becomes zero. Now, a question arises: what about points 
where $t_1$ (the variable we have associated to the dilaton) vanishes? 
  
However, before trying to answer this question, it is worth to point out that 
on the locus ${\cal C}_{1}=\{\phi^2=1\}$ we have, from (\ref{eq:160}),
\begin{equation}
t_1=0,
\end{equation}
so the candidate to look for an enhancement of 
gauge symmetry is the locus ${\cal C}_1$. We stress the fact that it is 
the $U(1)$ factor corresponding 
to the dilaton the one that becomes $SU(2)$.

Using the results stated in A.4 and A.6, we can qualitatively understand this 
enhancement of symmetry. To do so, we must first recall that the discriminant 
locus $\phi^2=1$ is a singular point for the Picard-Fuchs 
differential equation associated to the perestro\u{\i}ka corresponding to the 
resolution of the $A_1$ singularity. As explained in Appendix A.4, the 
preimage, 
in the resolution of this singularity, of the singular point is a $2$-cycle 
(there are $n-1$ $2$-cycles for a generic $A_{n-1}$ singularity). Now, we can 
consider $2$-branes wrapped around this $2$-cycle. The mass of these 
$2$-branes 
will become zero whenever the size of the $2$-cycle becomes zero. From 
equation 
(\ref{eq:160}) we observe that this is what precisely happens at the 
discriminant locus $\phi^2=1$ provided we measure the size of the $2$-cycle 
in terms of the corresponding K\"{a}hler form. At this point a subtlety 
arises  
that we would like to mention: in terms of the variable $z$, the one 
dimensional 
perestro\u{\i}ka defined by the resolution of the $A_1$ singularity is defined 
by going from $z=0$ to $z=\infty$ \cite{dis}. The discriminant locus 
corresponds to the 
singularity at $z=1$. Now, we parameterize this path in terms of $S$. From 
$z=0$ to $z=1$ we are moving along the the imaginary axis, taking from 
$S=i \infty$ to $S=0$; however, from $z=1$ to $z=\infty$, 
using (\ref{eq:160}), 
we get to the strange point $S=i0- \frac {1}{2}$.
   
This discussion leads to a natural question: we must now try to find out 
whether this non perturbative enhancement of symmetry has an equivalent 
when working at the level of the $(\hat{u},\epsilon)$-plane.
  
But before discussing this general issue let us first present in more 
concrete terms the Calabi-Yau interpretation of the 
$(\hat{u},\epsilon)$-plane. 
This interpretation is based on the following set of similarities:

\begin{itemize}
        \item[{i)}] The two conifold branches and the loci II and III play the 
        same roles.
        \item[{ii)}] The locus ${\cal C}_1=\{\epsilon^2=1\}$, 
        and the locus $\phi^2=1$.
        \item[{iii)}] Identification of the blow up of the weak coupling point 
        limit point $(\hat{u}=0,\tau= i \infty)$, and the blow up of the 
        conifold singularity $(x=1,y=0)$ leads to the relation 
        $\frac {1}{\tilde{u}^2}= \frac {yx^2}{(1-x)^2}$.
        \item[{iv)}] The $T$ transformation $(\hat{u},\tau) \rightarrow 
        (\hat{u},\tau+1)$ and the symmetry transformation 
        $A: \psi \rightarrow \alpha \psi$, $\phi \rightarrow -\phi$, with 
        $\alpha^{12}=1$.
        \item[{v)}] $A$ interchanges the conifold branches and the 
        ${\cal C}_1$ branches in the very same way as $T$.
        \item[{vi)}] The locus ${\cal C}_0=\{\psi=0\}$ in the $(x,\sqrt{y})$ 
        transforms under $A$ as shown in Figure~12,


\begin{figure}[htbp]
\centering

     \begin{picture}(400,170)

     \thicklines

     \put(150,20){\line(0,1){120}}  
     \put(100,80){\line(1,0){170}}

     \put(135,130){\line(1,0){15}} \put(135,30){\line(1,0){15}}
     \put(135,130){\vector(0,-1){40}} \put(135,30){\line(0,1){70}}

     \put(160,125){$\sqrt{y}$}
     \put(105,50){$A$}
     \put(250,45){$x$}
     \put(160,15){$\infty$}

     \end{picture}

\caption{{\em The $A$ transformation\/}.}

\end{figure}
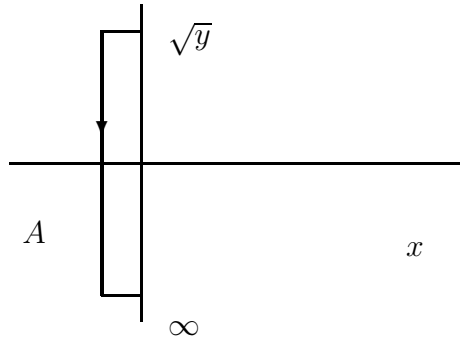


        which is, as Figure~13 clarifies, exactly what happens to the 
        locus I in the $(\hat{u},\epsilon)$-plane.


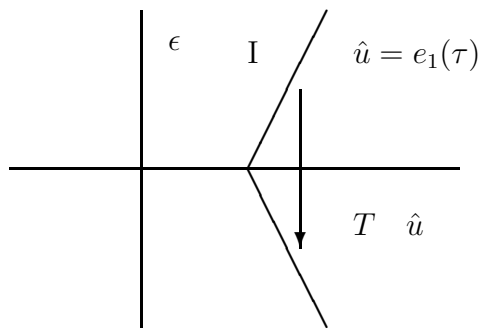
\begin{figure}[htbp]
\centering

     \begin{picture}(400,170)

     \thicklines

     \put(150,20){\line(0,1){120}}  
     \put(100,80){\line(1,0){170}}

     \put(190,80){\line(1,2){30}} \put(190,80){\line(1,-2){30}}
     \put(210,110){\vector(0,-1){60}} 

     \put(190,120){I} \put(230,120){$\hat{u}=e_1(\tau)$}
     \put(160,125){$\epsilon$} \put(230,55){$T$}
     \put(250,55){$\hat{u}$}
     
     \end{picture}

\caption{{\em The action of $T$ in the $(\hat{u},\epsilon)$-plane\/}.}

\end{figure}


        \item[{vii)}] In the Calabi-Yau case, $\phi \rightarrow -\phi$ can 
        be associated to the Weyl symmetry of an $A_1$ singularity.
\end{itemize}
  
All this similarities between the singular loci in the $(\hat{u},\tau)$-plane, 
and the moduli of complex structures of \IP$_{\{11226\}}$, can be connected 
through a one to one map between the two moduli spaces \cite{GHL},
\begin{equation}
{\displaystyle x=\frac {3/2e_{1}(\tau)}{3/2e_{1}(\tau)-\hat{u}},}
\hspace{1cm} {\displaystyle \sqrt{y}=-
\frac {e_{2}(\tau)-e_{3}(\tau)}{3e_{1} (\tau)},}
\label{eq:ex}
\end{equation}   
whose weak coupling limit reproduces relations (\ref{eq:149}),
\begin{equation}
{\displaystyle x =\frac {1}{1 \pm \hat{u}}+ \cdots} \hspace{1cm}
        y = \epsilon^2 + \cdots
\label{eq:le}
\end{equation}
  
Let us now finish with some general comments on the question raised above 
concerning the non perturbative enhancement of symmetry. The picture we have 
in mind can be summarized as follows: we start from the Donagi-Witten 
framework; the main point on that construction is the appearance of the 
$N\!=\!4$ moduli, $\tau$, used to measure the value of $\mbox{tr}\phi^2$ 
(for the $SU(2)$ case), and therefore to define the quantum moduli parameters 
of the $N\!=\!2$ gauge theory. Some sort of duality naturally appears in this 
context, between $\hat{u}$ and $\epsilon$: we can change the two dimensional 
Higgs field by varying the value of $\hat{u}$ for fixed $\epsilon$, or 
move the scale defined by the reference surface $E_{\tau}$ by moving its 
moduli, $\tau$. In both cases, and in the $(\tilde{u},\epsilon)$ variables 
used to make explicit the correspondence with the Calabi-Yau moduli space, 
the two descriptions are parameterized in terms of $\tilde{u}$. The Calabi-Yau 
interpretation we have proposed, relates the $N\!=\!4$ $\tau$ with the 
heterotic 
dilaton; in this way, the interchange between $\hat{u}$ and $\epsilon$ seems 
to be the reflection, at the level of the Donagi-Witten approach, of a duality 
symmetry between $S$ and $T$. In this spirit, the singularity loci defined 
by $\tau=0,1$ will appear as singularities of the $SU(2)$ theory described by 
fixing $\hat{u}$ and moving the moduli of the reference curve $E_{\tau}$. 
It is natural to conjecture that the richness of the Donagi-Witten model 
already contains the essence of heterotic-heterotic duality, codified 
in the dual role played in this scheme by $\hat{u}$ and $\tau$. This issue 
will be presented elsewhere.

\vspace{20 mm}
  
\begin{center}
{\bf Acknowledgments.}
\end{center}
  
It is a pleasure for C.G. to thank the organizers of the
conference, specially I. Todorov and A. Ganchev.
This work is partially supported by European Community grant 
ERBCHRXCT920069 and by grant PB92-1092. The work of R. Hernandez
is supported by U.A.M. fellowship. The work of E. L. is supported
by C.A.P.V. fellowship.

\newpage


\appendix

\section{Appendix.}

\subsection{Calabi-Yau Manifolds and Weighted Projective Spaces.}

The weighted projective space \IP$^{d+1}_{\{k_0 \ldots k_{d+1}\}}$, with 
homogeneous 
coordinates $[z_0,\ldots,z_{d+1}]$, is defined by the equivalence relation 
\begin{equation}
[z_0,\ldots,z_{d+1}] \sim [\lambda^{k_0}z_0,\ldots,\lambda^{k_{d+1}}z_{d+1}]
\end{equation}
A Calabi-Yau manifold of complex dimension $d$ can be defined as the vanishing 
locus of a homogeneous polynomial $W$, of degree $\sum_i k_i =k$:
\begin{equation}
W=\sum a_{i_0 i_1 \ldots i_{d+1}} z_0^{i_0} z_{1}^{i_1} \ldots 
z_{d+1}^{i_{d+1}}, 
\: \: \: \: \: \: \: \: \sum_{l=0}^{d+1}i_{l}k_l =k.
\label{eq:a2}
\end{equation}
Examples of vanishing polynomials are the following:
\begin{itemize}
        \item[{i)}] \IP$^{4}_{\{11222\}}$, \hspace{20 mm} $\sum k_i =8$.
        \begin{equation}
                W=z_1^8+z_2^8+z_3^4+z_4^4+z_5^4.
        \end{equation}
        \item[{ii)}] \IP$^4_{\{11226\}}$, \hspace{20 mm} $\sum k_i =12$.
        \begin{equation}
                W=z_1^{12}+z_2^{12}+z_3^6+z_4^6+z_5^2.
        \end{equation}
\end{itemize}
For some values of $a_{i_0 \ldots i_{d+1}}$ in (\ref{eq:a2}), 
for which $\frac {\partial W}{\partial z_i}=0$ has solution other than 
$z_i=0$ for all $z_i$, the manifold defined by $W$ develops singularities (see 
\ref{sec:msing} for more details on this point). 
For the above examples,
\begin{itemize}
        \item[{i)}] Singularity: $z_1=z_2=0$, $z_3^4+z_4^4+z_5^4=0$.
        \item[{ii)}] Singularity: $z_1=z_2=0$, $z_3^6+z_4^6+z_5^2=0$.
\end{itemize}
The values of $a_{i_0 \ldots i_{d+1}}$ for which the defined manifold is not 
smooth define the discriminant locus of the Calabi-Yau manifold; the 
discriminant 
locus is complex codimension one.
\begin{quote}
As an example, the discriminant locus of 
\begin{equation}
W=z_1^{12}+z_2^{12}+z_{3}^6+z_{4}^6+z_5^2-12 \psi z_1 z_2 z_3 z_4 z_5 -
2 \phi z_1^6 z_2^6
\end{equation}
is given by
\begin{equation}
\Delta= ((864 \psi^6 + \phi)^2 - 1)(\phi^2 -1).
\end{equation}
\end{quote}

\subsection{Toric Construction.}

Let $N \simeq {\bf Z}^n$ be the n-dimensional lattice of integer numbers. A 
fan $\Delta \subset N$ is defined as a set of cones $\sigma_i$ in the real
vector space $N_{\bf R} = N \otimes_{\bf R} {\bf R}$, such that the 
face of any cone $\sigma_i$ in $\Delta$ is also in $\Delta$.
The idea of toric geometry consists of associating to each cone a patch of 
coordinates, and to use the fan combinatorics to define the transition 
functions between different patches. In order to do so, some rules are needed:
\begin{itemize}
        \item[{R-1}] Given a cone $\sigma_i$, the dual cone $\check{\sigma}_i 
        \subset M_{\bf R}$, with $M$ the dual lattice to $N$, is defined 
        through
        \begin{equation}
                \check{\sigma}_i \cap M = Z_{\geq 0} m_{i,1} + 
                \cdots + Z_{\geq 0} m_{i,d_i}
        \end{equation}
        \item[{R-2}] The set (of vectors expanding the cones) $m_{i,l}$ 
        satisfies relationships of the form
        \begin{equation}
                \sum p_l^{(i)} m_{i,l} = 0.
        \end{equation}
        \item[{R-3}] To each $\sigma_{i}$ the patch of coordinates 
        $U_{\sigma_i}$ 
        is associated through
        \begin{equation}
        u_{\sigma_i}=\{ (u_{i,1}, \ldots, u_{i,d_i}): \prod_l 
        u_{i,l}^{p_l^{(i)}}=1 \}.
        \label{eq:a7}
        \end{equation}
        \item[{R-4}] Given the patches $U_{\sigma_i}$, $U_{\sigma_j}$, 
        in order to 
        find the coordinate change leading from one of them to the other, we 
        must find a relation of the form
        \begin{equation}
        \sum q_l^{(i)}m_{i,l}+ \sum q_{m}^{(j)} m_{j,m}=0
        \end{equation}\cite{B}
        \item[{R-5}] The transition functions are defined by
        \begin{equation}
        \prod_l u_{i,l}^{q_l^{(i)}} \prod_m u_{j,m}^{q_m^{(i)}} = 1.
        \end{equation}
\end{itemize}
  
As a simple example let us consider the case of a fan with only one cone, as 
that depicted in Figure~14. For the picture in $(a)$, the dual cone 
$\check{\sigma}$ 
is defined by vectors $(1,0)$ and $(-1,2)$, so that
\begin{equation}
\check{\sigma} \cap M = Z_{\geq 0} (1,0)+ Z_{\geq 0} (-1,2) + Z_{\geq 0} (0,1).
\end{equation}
Using (\ref{eq:a7}), we get
\begin{equation}
u_{(1,0)}^2=u_{(-1,2)}u_{(0,1)}.
\label{eq:a11}
\end{equation}


\begin{figure}[htbp]
\centering

        \begin{picture}(400,170)
                
        \put(30,60){\line(1,0){30}} \put(140,60){\line(1,0){30}} 
        \put(270,60){\line(1,0){30}}
        
        \thicklines

        \put(40,60){\vector(0,1){40}} \put(40,60){\vector(1,1){40}}
        \put(70,70){$(2,1)$} \put(10,105){$(0,1)$}
        \put(35,20){$(a)$} 

        \put(150,60){\vector(0,1){40}} \put(150,60){\vector(3,2){60}}
        \put(120,105){$(0,1)$} \put(205,70){$(3,1)$}
        \put(145,20){$(b)$}

        \put(280,60){\vector(0,1){40}} \put(280,60){\vector(2,1){80}}
        \put(250,105){$(0,1)$} \put(360,70){$(n,1)$}
        \put(275,20){$(c)$}

        \end{picture}

\caption{{\em Toric cones giving rise to singularities of different 
orders\/}.}

\end{figure}
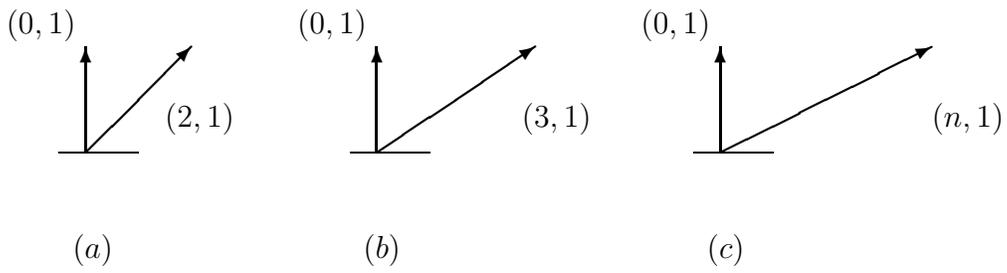


Defining $z \equiv u_{(1,0)}$, $x \equiv u_{(-1,2)}$ and $y \equiv u_{(0,1)}$, 
equation (\ref{eq:a11}) defines the manifold {\bf C}$^2/{\bf Z}_2$: $xy=z^2$. 
In just the same way, the cone in $(b)$ describes {\bf C}$^2/{\bf Z}_3$; in 
general, 
as in the diagram in $(c)$, we obtain the orbit space {\bf C}$^2/{\bf Z}_n$.
  
A Dynkin diagram $A_{n-1}$ can be associated to these spaces, defined as the 
set of internal lattice points on the edge $[(1,0),(n,1)]$.
  
\vspace{2 mm}
  
A more detailed example would be working out the manifold associated with the 
fan in Figure~15.


\begin{figure}[htbp]
\centering

        \begin{picture}(400,170)

        \thicklines

        \put(150,60){\vector(1,0){60}} \put(150,60){\vector(0,1){60}}     
        \put(150,60){\vector(-2,-1){70}}

        \put(200,40){$(1,0)$} \put(158,112){$(0,1)$} 
        \put(70,10){$(-2,-1)$}

        \put(202,100){$\sigma_1$} \put(160,20){$\sigma_2$} 
        \put(90,100){$\sigma_3$}

        \end{picture}

\caption{{\em Toric construction of the projective space \IP$_{\{112\}}$\/}.}

\end{figure}
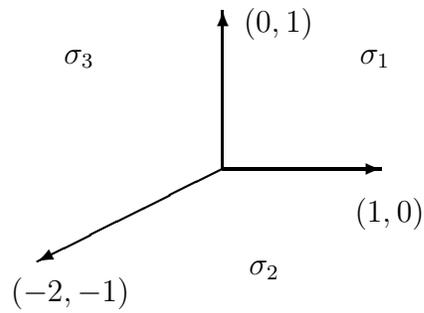


The dual cones $\check{\sigma}_i$ are given by 
\begin{eqnarray}
\check{\sigma}_1 & = & [(1,0),(0,1)], \nonumber \\
\check{\sigma}_2 & = & [(0,-1),(1,-2)], \nonumber \\
\check{\sigma}_3 & = & [(-1,0),(-1,2)].
\end{eqnarray}
Using the rules given above, coordinates in the different patches are given by 
\begin{eqnarray}
U_{\sigma_1} & = & (u,v), \nonumber \\
U_{\sigma_2} & = & (v^{-1},uv^{-2}), \nonumber \\
U_{\sigma_3} & = & (u^{-1},u^{-1}v^2,u^{-1}v).
\label{eq:a13}
\end{eqnarray}
It must be stressed that the chart $U_{\sigma_3}$ is again isomorphic to 
{\bf C}$^2$/{\bf Z}$_2$:
\begin{equation}
u^{-1}(u^{-1}v^2)=(u^{-1}v)^2.
\end{equation}
This is already clear from the cone $\sigma_3$, in the fan of Figure~15: 
the edge 
$[(0,1),(-2,-1)]$ contains one interior lattice point, $(-1,0)$. The space 
defined by (\ref{eq:a13}) is the weighted projective space \IP$_{\{112\}}$.


\subsection{Toric Blow up.}

The space {\bf C}$^2$/{\bf Z}$_n$ is an example of a singularity of 
type $A_{n-1}$, attending to Arnold's classification (see next paragraph). 
The resolution of this singularity consists in finding a smooth manifold 
${\cal V}$, and a proper map $\Pi : {\cal V} \rightarrow {\bf C}^2/{\bf Z}_n$, 
which is biholomorphic outside the preimage $\Pi^{-1}(0)$ of the 
singular point 
in {\bf C}$^2$/{\bf Z}$_n$. A toric approach to ${\cal V}$ is the construction 
of a new fan, obtained by adjoining vectors to the old one in such a way that 
all lattice points in each cone of the new fan are generated by the vectors 
defining the cone. 
  
As an example, consider the fan associated to 
{\bf C}$^2$/{\bf Z}$_2$ given in Figure~14.$(a)$. It is clear that in order to 
generate all lattice points in $\sigma$ we need to include the vector $(1,1)$. 
The new fan, defined by vectors $(0,1)$, $(1,1)$ and $(2,1)$, contains two 
patches, and defines the resolution of the singularity 
{\bf C}$^2$/{\bf Z}$_2$. 
  
From the above, it is clear that the blow up of the $A_{n-1}$ spaces 
{\bf C}$^2$/{\bf Z}$_n$ requires including as many extra vectors as 
points are there in the corresponding Dynkin diagram.


\subsection{Brief Review of Singularity Theory.}

\subsubsection{Singularities and Platonic Bodies.}

Let $\Gamma$ be a discrete subgroup of $SO(3)$, and let us denote by 
$\Gamma^*$ the preimage of $\Gamma$ by the covering map $SU(2) \rightarrow 
SO(3)$ (the classical notation for the cyclic, dihedral, tetrahedron, 
octahedron and icosahedron groups is, respectively, {\bf Z}$_n$, 
{\bf D}$_{2n}$, 
{\bf T}, {\bf O} and {\bf I}). The action of $\Gamma^*$ on {\bf C}$^2$ 
will be defined through 
the algebra ${\cal A}_{\Gamma^*}$ of polynomials in two complex variables 
invariant 
under the action of $\Gamma^*$. The generators of ${\cal A}_{\Gamma^*}$, 
denoted by $x$, $y$ and $z$, satisfy a relation $R_{\Gamma^*}(x,y,z)=0$, that 
defines a hypersurface ${\cal V}$ in {\bf C}$^3$.

\begin{center}
\begin{tabular}{|c|l|l|} \hline
        \multicolumn{1}{c}{$\Gamma^*$} & 
        \multicolumn{1}{c}{$R_{\Gamma^*}(x,y,z)=0$} & 
        \multicolumn{1}{c}{Dynkin diagram} \\   \hline
        {\bf Z}$_n$    & $xy=z^n$             & $A_{n-1}$  \\
        {\bf D}$_{2n}$ & $xy^2-x^{n+1}+z^2=0$ & $D_{n+2}$  \\
        {\bf T}        & $x^4+y^3+z^2=0$      & $E_6$      \\
        {\bf O}        & $x^3+xy^3+z^2=0$     & $E_7$      \\
        {\bf I}        & $x^5+y^3+z^2=0$      & $E_8$      \\ \hline
\end{tabular}
\end{center}

The hypersurface ${\cal V}$ defined by $R_{\Gamma^*}(x,y,z)=0$ is isomorphic 
to the orbit space {\cal C}$^2$/$\Gamma^*$. All these surfaces have a unique 
singular point ${\cal O}$. The resolution of the singularity takes place 
through 
a map $\Pi: \tilde{{\cal V}} \rightarrow {\cal V}$, where $\tilde{{\cal V}}$ 
is a smooth manifold; the preimage $\Pi^{-1}({\cal O})$ of the singular point 
is the union of 2-cycles $\Gamma_i$,
\begin{equation}
\Pi^{-1}({\cal O})=\Gamma_1 \cup \Gamma_2 \cup \ldots \cup \Gamma_{\mu},
\end{equation}
with $\mu$ the number of points of the corresponding Dynkin diagram. 
Furthermore, 
the intersection between these 2-cycles is determined by the intersection 
matrix of the Dynkin diagram. Thus, we associate 2-cycles $\Gamma_i$ with the 
points of the diagram, and a link $(i,j)$ whenever the corresponding 2-cycles 
$\Gamma_i$, $\Gamma_j$ intersect.

\subsubsection{Dynkin Diagrams and Picard-Lefschetz Theory.}

Let $f:{\bf C}^n \rightarrow {\bf C}$ be a Morse function\footnote{In the case 
considered in the text, that of Calabi-Yau threefolds, $n=3$.}, i.e., a f
unction 
such that all its critical points are non degenerate; the corresponding 
critical values will be denoted by $\alpha_1, \ldots, 
\alpha_{\mu}$. 
The level manifold ${\cal V}_{\alpha_i}$, associated to each critical value, 
is defined as follows:
\begin{equation}
{\cal V}_{\alpha_i}=\{(z_1,\ldots,z_n), \: f(z_1,\ldots,z_n)=\alpha_i\} \equiv 
f^{-1}(\alpha_i).
\end{equation}
Notice that ${\cal V}_{\alpha_i}$ are singular hypersurfaces of dimension 
$n-1$. 
The singular point $a_i \in {\cal V}_{\alpha_i}$ is the critical point with 
critical value $\alpha_i$.
  
Let us now consider a point $\alpha \in {\bf C}$, which corresponds to a 
regular 
value of $f$, so that the level manifold ${\cal V}_{\alpha}$ is non singular. 
From the regular point $\alpha$ to the critical values $\alpha_i$ a set of 
paths $\varphi_i$ can be defined, through
\begin{equation}
\varphi_i(\tau)=\left\{ \begin{array}{ll}   \alpha & \tau=0 \\
                                         \alpha_i & \tau=1, \end{array}
                                         \right.
\end{equation}
and such that $\varphi_i(\tau)$ is a regular value of $f$ for $\tau \neq 1$. 
  
Given the homology $H_{n-1}({\cal V}_{\alpha})$, a vanishing cycle 
$\Delta_i \in H_{n-1}({\cal V}_{\alpha})$ will be an $(n-1)$-cycle that, when 
``transported'' by the path $\varphi_i$ will contract to the critical point 
$a_i$, i. e., to the singular point of the level manifold 
${\cal V}_{\alpha_i}$; 
therefore, a vanishing cycle $\Delta_i$ can be associated to each path 
$\varphi_i$. They define a basis of $H_{n-1}({\cal V}_{\alpha})$. We can now 
construct a Dynkin diagram, associated to $f$, by putting the points of the 
diagram in a one to one correspondence with the vanishing cycles 
$\Delta_i$ $(i=1, \ldots , \mu)$; the intersection matrix of the diagram 
will be the intersection form in $H_{n-1}({\cal V}_{\alpha})$.
  
For each path $\varphi_i$ we can define a loop $\gamma_i$ starting at the 
point $\alpha$, and going around the critical value $\alpha_i$. 
Picard-Lefschetz theory associates to each loop $\gamma_i$ a monodromy 
matrix $h_i$,
\begin{equation}
h_i: H_{n-1}({\cal V}_{\alpha}) \rightarrow  H_{n-1}({\cal V}_{\alpha})
\end{equation}

As an example, consider the Landau-Ginzburg potential
\begin{equation}
W=z^n + t_1 z.
\end{equation}
In the $w$-plane we have $(n-1)$-critical values $\alpha_i$. Let us denote by 
$a_i$ the corresponding critical points, i. e., the different vacua of $W$. 
The 
level manifold ${\cal V}_{\alpha}$ at a regular point $\alpha$ consist of the 
set of $n$ points 
\begin{equation}
{\cal V}_{\alpha} = \{ z^n+t_1z = \alpha \}.
\end{equation}
Calling these points $z_1, \ldots ,z_n$, the homology 
$H_{0}({\cal V}_{\alpha})$ 
is generated by the $n-1$ $0$-cycles $\Delta_i=[z_i,z_{i+1}]$, 
$i=1,\ldots,n-1$. 
now, we define the set of paths $\varphi_i$ in the $w$-plane (see Figure~16). 
When we 
transport the cycle $\Delta_i$ from $\alpha$ to $\alpha_i$, we immediately 
observe that this $0$-cycle contracts to the critical point $a_i$. Therefore, 
we have $n-1$ vanishing cycles in $H_{0}({\cal V}_{\alpha})$. 
It is then immediate 
to see that they define the Dynkin diagram $A_{n-1}$.


\begin{figure}[htbp]
\centering

     \begin{picture}(400,170)

     \thicklines

     \put(150,20){\line(0,1){100}}  
     \put(120,40){\line(1,0){150}}
     
     \put(80,45){\circle*{3}} \put(70,32){$\alpha$} 
     \put(95,32){$\varphi_1$}
     \put(140,45){\circle*{3}} \put(130,32){$\alpha_1$} 
     \put(160,20){\circle*{3}} \put(165,15){$\alpha_2$}
     \put(185,30){\circle*{3}} \put(190,25){$\alpha_3$}
     \put(190,60){\circle*{3}} \put(195,55){$\alpha_4$}
     \put(210,80){\circle*{3}} \put(215,75){$\alpha_5$}
     \put(180,95){\circle*{3}} \put(185,105){$\alpha_{n-1}$}
     \put(130,88){$\varphi_{n-1}$}

     \put(80,45){\line(1,0){60}} \put(80,45){\line(2,1){100}}

     \put(160,125){$w$-plane}
     
     \end{picture}

\caption{{\em Paths in the $w$-plane used in the definition of the homology 
basis\/}.}

\end{figure}
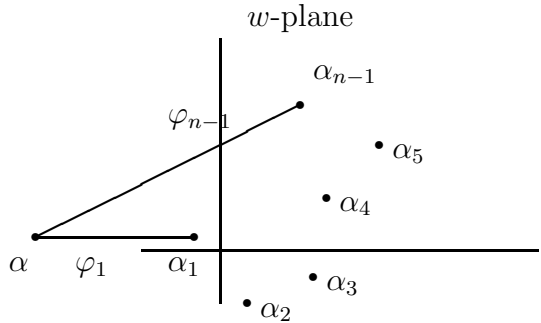



\subsection{Discriminants and Perestro\u{\i}kas.}
\label{sec:msing}

Let us consider the Calabi-Yau manifold $X$ in \IP$_{\{11226\}}$ defined by 
\begin{equation}
x_1^{12}+x_2^{12}+x_3^{6}+x_4^{6}+x_5^{2}=0.
\label{eq:p1}
\end{equation}
To study the K\"{a}hler classes of (\ref{eq:p1}), we can consider the complex 
deformations of its mirror manifold $Y$ defined by
\begin{equation}
x_1^{12}+ x_2^{12}+ x_3^{6}+ x_4^{6}+ x_5^{2}+
{\tilde a_1} a_1x_1^6x_2^6+ {\tilde a_2} a_2x_1x_2x_3x_4x_5=0.
\label{eq:p2}
\end{equation}
When noticing that all terms in (\ref{eq:p2}) are of the form $x_1^{n_1}
x_2^{n_2} x_3^{n_3} x_4^{n_4} x_5^{n_5}$, a point in {\bf R}$^5$ can be 
associated to each monomial: $(n_1,n_2,n_3,n_4,n_5)$. Therefore, equation 
(\ref{eq:p2}) is represented by a collection of (seven) points in {\bf R}$^5$. 
Following notation from reference \cite{dis} we will denote this set by 
${\cal A}$:
\begin{eqnarray}
{\cal A} & \equiv & \{(12,0,0,0,0),(0,12,0,0,0),(0,0,6,0,0),(0,0,0,6,0), 
\nonumber \\
         &        & (0,0,0,0,2),(6,6,0,0,0),(1,1,1,1,1)\}.
\end{eqnarray}
The set ${\cal A}$ lies in a four dimensional polytope $P$, called 
Newton polytope, whose corners are determined by the vectors associated 
to the monomials in (\ref{eq:p1}).
   
Defining polynomials for faces $\Gamma$ in the Newton polytope can be easily 
done by just considering the set of monomials associated to points in
that face. The 1-dimensional face spanned by vectors $(12,0,0,0,0),
(0,12,0,0,0)$ is the polytope corresponding to 
\begin{equation}
W_{\Gamma}=x_1^{12}+x_2^{12}+{\tilde a_1}x_1^6x_2^6.
\label{eq:wg}
\end{equation}

This decomposition into faces of the Newton polytope also provides a 
powerful tool for the characterization of the singularities arising in the
moduli space. The number of points in $\cal A$ that lie in 
the interior of the face determined by the monomials $x_{1}^{12}$ and
$x_{2}^{12}$, namely one point: $(6,6,0,0,0)$, implies the 
existence of a singularity of $A_1$ type \cite{B}.
 
\vspace{2 mm}
  
Let us consider now a triangulation of the set ${\cal A}$.
This triangulation will define a fan of cones, in terms of which we 
can give a toric representation of the manifold (\ref{eq:p2}). 
Part of the idea of topology changing amplitudes is related to the
existence of different triangulations for the same set ${\cal A}$ of 
points.
    
\vspace{2 mm}
  
Before entering the question of topology change, we will concentrate on 
the discriminant $\Delta$ of (\ref{eq:p2}). 
We consider a homogeneous polynomial 
\begin{equation}
W=\sum a_i m_i
\end{equation}
defined in terms of the set of monomials $m_i=\prod x_l^{n_l}$, 
where we have introduced parameters $a_i$ for each monomial.
Then, the discriminant 
is the condition obtained on the coefficients $a_i$ by requiring that there 
exist solutions to 
\begin{equation}
\frac {\partial W}{\partial x_i}=0, 
\end{equation}
for $x_i$ not equal zero for all $i$. Partial discriminants (those coming 
from the corresponding Landau-Ginzburg 
potential $W_{\Gamma}$) $\Delta_{\Gamma}(a_i)$ are naturally associated to 
faces $\Gamma$ of the Newton polytope. In 
this way, the face $\Gamma$ determined by vectors 
$(12,0,0,0,0),(0,12,0,0,0)$, whose 
defining polynomial is that given in 
(\ref{eq:wg}),
leads to the discriminant 
\begin{equation}
\Delta_{\Gamma}({\tilde a_1})=(1 - \frac {4}{{\tilde a_1}^2}).
\label{eq:p8b}
\end{equation}
  
Vanishing of the discriminant defines the discriminant locus, the point in 
moduli space (parameterized by $a_i$) where the manifold defined as the locus 
$W=0$ becomes singular.
  
\vspace{2 mm}
  
Now we can try to understand the discriminant locus, in string language,
from the type II$_A$ or II$_B$ point of view. In order to do so, we
will make use of mirror symmetry to interpret the quantities $a_i$
parameterizing the complex structures of the manifold $Y$, defined by $W$,
in terms of K\"ahler classes of its mirror $X$.

In example (\ref{eq:p2}), the moduli space of complex structures,
parameterized by ${\tilde a_1}$ and 
${\tilde a_2}$, is isomorphic to $({\bf C}^*)^2$ 
$\hspace{2mm}$\footnote{By rescaling 
$x_i \rightarrow \lambda_i x_i$, five of the possible 
deformations $a_i$ have been set to one. The (reduced) moduli space is 
\[ {\cal M} \simeq \frac {({\bf C}^*)^7}{({\bf C}^*)^5} \simeq 
{({\bf C}^*)^2}. \]
See \cite{dis} for a detailed discussion on this point.}.
Let us now define new coordinates $u_i$ by 
\begin{equation}
u_i=- \frac {1}{2\pi} \log | \tilde{a_i} |
\label{eq:p9}
\end{equation}
in {\bf R}$^2$. These coordinates can be interpreted in the spirit of the 
monomial divisor map \cite{mon}. In fact, the K\"{a}hler classes in $H_2(X)$ 
can be parameterized in terms of quantities $t_i$ related to the complex 
deformation variables $\tilde{a}_i$ by 
$\tilde{a}_i = e^{2 \pi i t_i}$. Asymptotically (${\tilde a}_i 
\rightarrow \infty$), the 
imaginary part of $t_i$, given by (\ref{eq:p9}), is the (real) K\"{a}hler 
form $J_i$ in (4.21). 
  
The structure of K\"{a}hler cones on {\bf R}$^2$ can be now determined 
by using 
the properties of the discriminant: when moving to large values of 
$\tilde{a}_i$, 
we notice that the discriminant is dominated by some particular monomial; we 
will refer to this monomial by $r_{\Delta_i}$. Therefore, {\bf R}$^2$ is 
divided into sectors where a particular monomial $r_{\Delta_i}$ is dominating, 
as indicated in Figure~17.


\begin{figure}[htbp]
\centering

     \begin{picture}(400,170)

     \thicklines

     \put(150,40){\line(-1,1){60}}  
     \put(150,40){\line(1,2){40}}
     \put(150,40){\line(3,-1){100}}
     
     \put(150,40){\circle*{3}} \put(140,90){$r_{\Delta_1}$} 
     \put(200,60){$r_{\Delta_2}$}

     \put(200,100){\vector(-1,0){20}} \put(205,98){wall}
     \put(60,60){{\bf R}$^2$}

     \end{picture}

\caption{{\em K\"{a}hler cones structure as arising from the discriminant\/}.}

\end{figure}
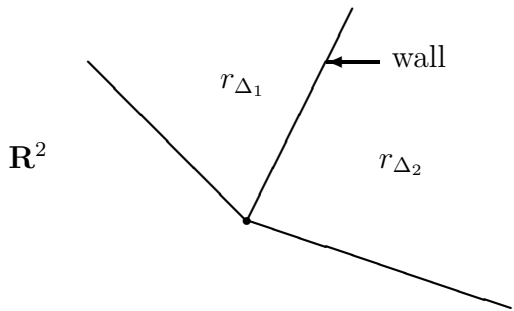

The discriminant locus now appears concentrated on the {\em walls\/}  
separating different regions.
  
A beautiful result in toric geometry allows us to associate with each monomial 
$r_{\Delta_i}$ a particular triangulation of the set of points ${\cal A}$ 
determining the Newton polytope: given a triangulation $\Delta_i$ (just as 
the one the monomials dominating in the discriminant along different regions 
introduce) of ${\cal A}$, then
\begin{equation}
r_{\Delta_i} = \prod a_k^{\sum_{\sigma \in \Delta_i, \: \alpha_k \in \sigma}
\mbox{vol}(\sigma)},
\end{equation}
where the sum is over all triangles with $\alpha_k$ (the vector defined by 
the monomial in $W$ with $a_k$ coefficient) as a vertex.
  
With the above geometry, whenever a wall separating different K\"{a}hler 
cones is transversed, a transition between different triangulations 
takes place; this process is called a {\it perestro\u{\i}ka}.
This transition represents a change in 
topology\footnote{The topology 
change does not modify the Hodge numbers, but the rational curves lying 
on the cones 
the wall bounds.}, so it is a candidate to topology changing amplitudes. 
In the particular case given by (\ref{eq:p2}), we can expect transitions 
on the wall  
determined by vanishing of the discriminant locus $
\Delta_{\Gamma}({\tilde a_1})$, defined 
in (\ref{eq:wg}).
  

\subsubsection{Picard-Fuchs Equations.}

To each point in the moduli space of complex structures, parameterized by 
$a_i$, we can associate a fiber defined by the $H_2$ cohomology of its 
mirror manifold $X$; elements in $H_2$ of the mirror will be denoted by  
$\Phi(a_i)$. Armed with this fibration, we can understand what 
happens as we move around the discriminant locus (This is exactly the same 
problem we have delt with in previous section, in connection with singularity 
theory; in that case, we had the homology $H_{n-1}({\cal V}_{\alpha})$ for 
the map $f: {\bf C}^n \rightarrow {\bf C}$, and we can study using  
Picard-Lefschetz theory the monodromy matrix arising in 
$H_{n-1}({\cal V}_{\alpha})$ 
when moving around critical values, which are just the equivalent notion, 
within this context, to the discriminant locus.).
  
The first thing we need to do is defining coordinates associated to 
each particular triangulation of the set $\cal A$; then, a differential 
Picard-Fuchs equation will be associated to transitions between different 
triangulations. In order to interpret the complete moduli space of
complex structures of $Y$ in terms of K\"ahler classes,
it is neccessary to include triangulations that can omit points in $\cal A$
while always containing the Newton polytope $P$ \cite{dis}.

In the example we are dealing with a simple perestro\u{\i}ka takes place. 
Defining
\begin{eqnarray} 
\alpha_1 & = & (6,6,0,0,0), \nonumber \\
\alpha_2 & = & (12,0,0,0,0), \nonumber \\
\alpha_3 & = & (0,12,0,0,0), 
\end{eqnarray}
the only two possible triangulations are those depicted in Figure 18.


\begin{figure}[htbp]
\centering

     \begin{picture}(400,170)

     \thicklines

     \put(100,40){\line(0,1){80}}  
     \put(80,38){$\alpha_3$} \put(80,118){$\alpha_2$}
     \put(100,40){\circle*{3}} \put(100,120){\circle*{3}}
     \put(96,10){$(a)$}

     \put(220,40){\line(0,1){80}}
     \put(200,38){$\alpha_3$} \put(200,118){$\alpha_2$}
     \put(200,80){$\alpha_1$} 
     \put(220,40){\circle*{3}} \put(220,80){\circle*{3}}
     \put(220,120){\circle*{3}}
     \put(216,10){$(b)$}

     \end{picture}

\caption{{\em The only possible Perestro\u{\i}ka in one dimension\/}.}

\end{figure}
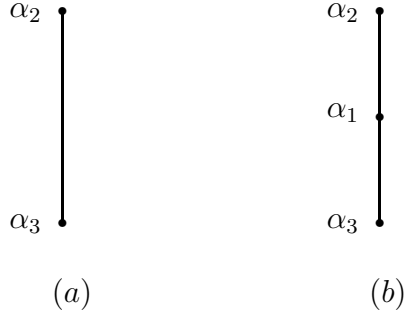

The triangulation Figure 18.$(b)$ corresponds to resolution of the $A_1$ 
singularity, as 
we have included the vertex $\alpha_1$; in this situation, we have a cone 
in the 
fan of the manifold with one edge going from the origin to the point 
$\alpha_1$. 
On the other hand, the triangulation in $(a)$ does not see the point 
$\alpha_1$, so that it produces a singular space, {\bf C}$^2$/{\bf Z}$_2$.

\vspace{2 mm}
  
Following \cite{dis}, a recipe can be given to build up the differential 
equation associated to a 
simple perestro\u{\i}kas between sets of points where only two triangulations 
are possible:
Perestro\u{\i}kas are associated with sets of $N+2$ points in {\bf R}$^N$ that 
are not contained in an {\bf R}$^{N-1}$ hyperplane. Denoting by $\alpha_l$ 
these 
points, we must find relations of the form
\begin{equation}
\sum m_l \alpha_l = 0 
\end{equation}
in order to define the invariant coordinate
\begin{equation}
z \equiv \prod a_l^{m_l}.
\end{equation}
The differential operator is defined as
\begin{equation}
\Box = \prod_{m_l>0} \left( \frac {\partial}{\partial a_l} \right) ^{m_l} - 
\prod_{m_l<0} \left( \frac {\partial}{\partial a_l} \right) ^{-m_l}  ,
\label{eq:p15}
\end{equation}
where $a_l$ stands for the coefficients of the monomials in the polynomial $W$.

For the example in Figure 18, we get\footnote{For convenience, the factor 4 
(appearing in expression (\ref{eq:p8b}) for the discriminant 
$\Delta_{\Gamma}(a_1)$) 
has been introduced.}
\begin{equation}
2\alpha_1 = \alpha_2 +\alpha_3,
\label{eq:p16a}
\end{equation}
so that
\begin{equation}
z= 4 \frac {a_2 a_3}{a_1^2}.
\label{eq:p16}
\end{equation}
As in (\ref{eq:p2}) we have chosen $a_2=a_3=1$, $z$ becomes 
$z = \frac {4}{{\tilde a_1}^2}$, 
and the operator (\ref{eq:p15}) is
\begin{equation}
\Box = \left( z \frac {d}{dz} \right) ^2 - z \left( z \frac {d}{dz} \right)
\left( z \frac {d}{dz} + \frac {1}{2} \right).
\end{equation}
The invariant $z$ parameterizes the transition between the two triangulations. 
  
From (\ref{eq:p8b}) we notice that the discriminant is given by $z=1$. 
Moreover, 
the solution obtained from $\Box f=0$,
\begin{equation}
f= C_1 + C_2 \log \left( \frac {2-z-2\sqrt{1-z}}{z} \right),
\label{eq:p18}
\end{equation}
has non trivial monodromy around $z=1$ (namely, $-\II$).
  
Through the monomial divisor map, equation (\ref{eq:p18}) provides us 
with the coordinate 
in $H_2(X)$ corresponding to $z$:
\begin{equation}
t= \frac {1}{2\pi i} \log \left( \frac {2-z-2\sqrt{1-z}}{z} \right).
\label{eq:p19}
\end{equation}
The following facts, arising from (\ref{eq:p19}), must be stressed:
\begin{itemize}
        \item[{i)}] $t=0$ when $z$ is in the discriminant locus $(z=1)$.
        \item[{ii)}] As we move around the discriminant locus, we pick up 
        a monodromy for t which is the Weyl group of the $A_1$ singularity 
        (\ref{eq:p16a}).
        \item[{iii)}] Moving from $z=1$ to $z=0$ the singularity in 
        {\bf C}$^2$/{\bf Z}$_2$ has been blown up to infinite size.
        \item[{iv)}] An interesting point is the size of the $2$-cycle used 
        to blow up the $A_1$ singularity: the point $z=\infty$ is expected to 
        correspond to a singular space and, certainly, $J=0$; however, $B$ is 
        not zero at that point ($B=-\frac{1}{2}$). On the other hand, at 
        the discriminant locus $z=1$, we have $B+iJ=0$, the difference being 
        the value of the $B$ field.
\end{itemize}

\newpage


\end{document}